\documentclass[10pt,aps,prd,showpacs,twocolumn,superscriptaddress,nofootinbib,floatfix]{revtex4-1}
\usepackage{amsfonts,amsmath,amssymb,amsthm}
\usepackage[bookmarks, bookmarksopen, bookmarksnumbered]{hyperref}
\usepackage{graphicx}
\usepackage{tensor}
\usepackage{float}
\usepackage{color}
\usepackage{epigraph}

\begin{document}

\title{Toroidal magnetic fields in self-gravitating disks around black holes}
\author{Wojciech Dyba}
\author{Patryk Mach}
\author{Miko{\l}aj Pietrzy\'{n}ski}
\affiliation{Instytut Fizyki Teoretycznej, Uniwersytet Jagiello\'{n}ski, \L ojasiewicza 11,  30-348 Krak\'ow, Poland}

\begin{abstract}
We investigate stationary models of magnetized, self-gravitating disks around black holes. The disks are assumed to rotate according to a recently introduced Keplerian rotation law. We consider different prescriptions of the toroidal magnetic field. Similarly to the purely hydrodynamical case (i.e., with no magnetic field), we observe a bifurcation in the parameter space of solutions. There are usually two branches of solutions: a branch corresponding to relatively light disks and a branch for which the disk can be more massive than the black hole. The existence of this latter branch can be explained by geometric properties of the spacetime. We investigate the influence of the magnetic field in the disk on these effects.
\end{abstract}

\maketitle

\section{Introduction}

The existing general-relativistic theory of stationary self-gravitating magnetized fluids has been developed mostly in the context of rotating stars. In seminal papers \cite{bekenstein_oron_const,bekenstein_oron} Bekenstein and Oron developed a formalism which should in principle allow one to dress up a suitable non-magnetized model with poloidal magnetic fields. Their construction assumed a particular form of the toroidal electric current within the star. More general configurations were subsequently studied in \cite{bonazzolla1993,bocquet1995,konno1999,cardall2001,zanotti2002,ioka2003,ioka2004,gorg}. In particular Ref.\ \cite{gorg} contains a very general (an also slightly formal) analysis of possible stationary configurations. They can be classified with respect to the morphology of magnetic fields (poloidal, toroidal, or both) and the fluid velocity (meridional flow, differential rotation).

Much less is known about self-gravitating, magnetized toroidal configurations. A Newtonian model of this kind was constructed by Otani, Takahashi, and Eriguchi in \cite{otani_newtonian}. In Ref.\ \cite{zanotti_pugliese} Zanotti and Pugliese have analyzed the so-called von Zeipel property in the general context of magnetized disks, both in the Newtonian and the general-relativistic case. In Ref.\ \cite{magnetic_nasze} Mach, Gimeno-Soler, Font, Odrzywo{\l}ek, and Pir\'{o}g, obtained numerical solutions representing stationary self-gravitating magnetized perfect fluid tori around black holes.

The literature on test-fluid (i.e., non self-gravitating) stationary disks with magnetic fields is more numerous. One of the first exact solutions was derived by Komissarov \cite{komissarov}. Other works in this spirit include \cite{montero, soler, witzany, lahiri, cruzosorio, faraji1, faraji2}. The authors of Refs.\ \cite{schroven,trova} adopted a framework opposite to ideal general-relativistic magnetohydrodynamics and considered a case of charged disks with a vanishing conductivity.

This paper is a sequel of Refs.\ \cite{magnetic_nasze} and \cite{ergosfery}. In the latter, we investigated purely hydrodynamical models (i.e. with no magnetic fields), focusing on strong gravitational field effects connected with massive tori. In both papers we have assumed a Keplerian rotation prescription proposed in Refs.\ \cite{kkmmop,kkmmop2}. One of striking observations made in \cite{ergosfery} is an occurrence of a bifurcation in the parameter space of solutions. Roughly speaking, for fixed black-hole mass and spin, the maximal rest-mass density within the torus and geometric parameters: inner and outer radii of the torus, there usually exist two solutions differing in the total mass of the system (or, equivalently, the mass of the torus). This is a purely relativistic effect, and it is related to the change of the torus volume. Another effect observed for massive disks is the occurrence of toroidal ergoregions, which can exist in addition to the ergoregion associated with the rotating black hole. Such toroidal ergoregions can appear inside the tori, but it is also possible to obtain configurations with the torus encompassed by the ergoregion. For sufficiently massive and compact systems, the two ergoregions can merge, forming a connected region bounded by the ergosurface of a spherical topology.

The purpose of the analysis presented in this paper is twofold. On one hand, we would like to explore the parameter space of solutions representing moderately massive magnetized disks around spinning black holes, focusing on parameters controlling the distribution of the magnetic field. On the other hand, we check how the magnetic field affects strong-gravitational field effects discovered in \cite{ergosfery}, in particular bifurcation diagrams.

A numerical scheme used in this paper was originally proposed in \cite{shibata}, and developed slightly in \cite{kkmmop,kkmmop2}. Magnetic terms have been added in \cite{magnetic_nasze}. We use a slightly improved version of the numerical code described in Ref.\ \cite{magnetic_nasze}.

We use standard gravitational system of units with $c = G = 1$, where $c$ is the speed of light, and $G$ is the gravitational constant. Greek indices will be used to label spacetime dimensions $\mu = 0,1,2,3$. Latin indices $i = 1,2,3$ will be reserved for spatial dimensions. 

\section{Stationary self-gravitating disks with toroidal magnetic fields}

\subsection{Euler-Bernoulli equation}

We work in the standard framework of the ideal General-Relativistic Magnetohydrodynamics (GRMHD). The energy-momentum tensor $T^{\mu \nu} = T_\mathrm{FLUID}^{\mu \nu} + T_\mathrm{EM}^{\mu \nu}$ is a sum of two components: the energy-momentum tensor of the perfect fluid
\[ T_\mathrm{FLUID}^{\mu \nu} = \rho h u^\mu u^\nu + p g^{\mu \nu} \]
and the energy-momentum tensor of the electromagnetic field
\[ T_\mathrm{EM}^{\mu \nu} = F^{\mu \alpha} F\indices{^\nu_\alpha} - \frac{1}{4} g^{\mu \nu} F^{\alpha \beta} F_{\alpha \beta}. \]
Here $\rho$ is the rest-mass density, $p$ is the thermal pressure, $u^\mu$ denotes the four-velocity of the fluid, $g_{\mu \nu}$ is the metric tensor, and $F^{\mu \nu}$ denotes the electromagnetic (Faraday) tensor.

The dual of the Faraday tensor is assumed in the form $^\ast F^{\mu\nu} = b^\mu u^\nu - b^\nu u^\mu$, where $b^\mu$ is the so-called four vector of the magnetic field. We assume that
\begin{equation}
\label{orthogonal}
b_\mu u^\mu = 0.
\end{equation}
This implies that $\nabla_\mu {}^\ast F^{\mu \nu} = 0$. In terms of the four-velocity $u^\mu$ and the four-vector $b^\mu$, the tensor $T_\mathrm{EM}^{\mu \nu}$ can be expressed as
\[ T_\mathrm{EM}^{\mu \nu} = \left( u^\mu u^\nu + \frac{1}{2} g^{\mu \nu} \right) b^2 - b^\mu b^\nu, \]
where $b^2 = b_\mu b^\mu$.
The total energy-momentum tensor reads
\begin{equation}
\label{emtensor}
T_{\mu \nu} = (\rho h + b^2) u_\mu u_\nu + \left( p + \frac{1}{2} b^2 \right) g_{\mu \nu} - b_\mu b_\nu.
\end{equation}
Note that the quantity $p_\mathrm{mag} = \frac{1}{2} b^2$ plays the role of a magnetic pressure.

We assume the metric of the form
\begin{equation}
\label{generalmetric}
g = g_{tt} dt^2 + 2 g_{t \varphi} dt d\varphi + g_{rr} dr^2 + g_{\theta \theta} d\theta^2 + g_{\varphi \varphi} d\varphi^2, 
\end{equation}
where $(t, r , \theta, \varphi)$ are spherical coordinates, and functions $g_{tt}$, $g_{t\varphi}$, $g_{rr}$, $g_{\theta \theta}$, $g_{\varphi \varphi}$ depend only on $r$ and $\theta$. Later, we will also specialize to the following quasi-isotropic form
\begin{eqnarray}
\nonumber
g & = & - \alpha^2 dt^2 + \psi^4 e^{2q} (dr^2 + r^2 d\theta^2) + \\
&& \psi^4 r^2 \sin^2 \theta (\beta dt + d \varphi)^2.
\label{isotropic}
\end{eqnarray}
Both forms of the metric admit two Killing vectors with contravariant components $\eta^\mu = (0,0,0,1)$ and $\xi^\mu = (1,0,0,0)$.

We assume that both the four-velocity and the four-vector of the magnetic fields are purely toroidal, i.e., $u^r = u^\theta = b^r = b^\theta = 0$. It can be easily shown that
\begin{equation}
b_t = - \frac{u^\varphi}{u^t} b_\varphi = - \Omega b_\varphi,
\end{equation} 
where $\Omega = u^\varphi/u^t$, and
\begin{equation}
b_\varphi^2 = - (u^t)^2 \mathcal L b^2,
\end{equation}
where $\mathcal L = g_{\varphi \varphi} g_{tt} - g_{t \varphi}^2$. The component $u^t$ can be expressed in terms of $\Omega$ as
\begin{equation}
\label{ut}
g_{tt} + 2 g_{t \varphi} \Omega + g_{\varphi \varphi} \Omega^2 = - \frac{1}{(u^t)^2}.
\end{equation}

The conservation equations
\begin{equation}
\label{cons_eqs_general}
\nabla_\mu \left( \rho u^\mu \right) = 0, \quad \nabla_\mu T^{\mu \nu} = 0
\end{equation}
can be integrated, yielding
\begin{equation}
\label{bernoulli3}
\int j(\Omega) d\Omega + \ln \left( \frac{h}{u^t} \right) + \int \frac{d(b^2 \mathcal L)}{2 \rho h \mathcal L} = C,
\end{equation}
where $C$ denotes an integration constant, and where we have assumed that the angular momentum per unit inertial mass, $j = u^t u_\varphi$, is a function of $\Omega$ only. Note that in the case with no magnetic fields, this is actually an integrability condition for Eqs.\ (\ref{cons_eqs_general}). Similarly, the quantity $f \equiv b^2 \mathcal L$ must be a function of $x \equiv \rho h \mathcal L$.

The electric current four-vector $\mathcal J^\nu = \nabla_\mu F^{\mu \nu}$ can be computed a posteriori. We have $\mathcal J^t = 0$, $\mathcal J^\varphi = 0$, but $\mathcal J^r \not\equiv 0$ and $\mathcal J^\theta \not\equiv 0$. In general,
\begin{equation}
\label{electricc}
\mathcal J^\nu = \rho_q u^\nu + \sigma F^{\nu \mu} u_\mu,
\end{equation}
where $\rho_q$ is the charge density, and $\sigma$ denotes the electrical conductivity. Since, in our case $u^r = 0$ and $u^\theta = 0$, we conclude that only the second term in Eq.\ (\ref{electricc}) contributes to the electric current. Note, that the ideal general-relativistic magnetohydrodynamics used in this paper corresponds to $\sigma \to \infty$ and $F^{\nu \mu} u_\mu \to 0$.

\subsection{Einstein equations}

Our formulation of Einstein equations, described in \cite{magnetic_nasze}, follows closely the construction of Shibata \cite{shibata}. Shibata's scheme guarantees that in the absence of the torus, i.e., for the vanishing rest-mass density, the pressure, and the terms related to magnetic fields, the solution tends explicitly to the Kerr metric. The Kerr solution can be written in the quasi-isotropic coordinates (\ref{isotropic}) as \cite{shibata,brandtseidel}
\begin{eqnarray}
g & = & - \alpha_\mathrm{K}^2 dt^2 + \psi_\mathrm{K}^4 e^{2q_\mathrm{K}} (dr^2 + r^2 d\theta^2) + \nonumber \\
&& \psi_\mathrm{K}^4 r^2 \sin^2 \theta (\beta_\mathrm{K} dt + d \varphi)^2,
\end{eqnarray}
where
\begin{subequations}
\begin{eqnarray}
\psi_\mathrm{K} & = & \frac{1}{\sqrt{r}}\Bigl( r^2_\mathrm{K}  +a^2 +2ma^2\frac{r_\mathrm{K}\sin^2\theta  }{\Sigma_\mathrm{K}}\Bigr)^{1/4}, \\
\beta_\mathrm{K} & = & -\frac{2mar_\mathrm{K}}{(r^2_\mathrm{K}+a^2)\Sigma_\mathrm{K} +2ma^2r_\mathrm{K} \sin^2\theta}, \\
\alpha_\mathrm{K} & = & \left[ \frac{ \Sigma_\mathrm{K} \Delta_\mathrm{K}}{(r_\mathrm{K}^2+a^2)\Sigma_\mathrm{K}+2ma^2r_\mathrm{K} \sin^2\theta} \right]^{1/2}, \label{eq:alpha_K}\\
e^{q_\mathrm{K}} & = & \frac{\Sigma_\mathrm{K}}{\sqrt{(r^2_\mathrm{K}+a^2)\Sigma_\mathrm{K} +2ma^2r_\mathrm{K} \sin^2\theta}},
\end{eqnarray}
\end{subequations}
and where we have defined
\begin{subequations}
\begin{eqnarray}
r_\mathrm{K} & = & r \left( 1 + \frac{m}{r} + \frac{m^2 - a^2}{4 r^2} \right), \\ 
\Delta_\mathrm{K} & = & r_\mathrm{K}^2 -2r_\mathrm{K}+a^2, \\
\Sigma_\mathrm{K} & = & r_\mathrm{K}^2 + a^2 \cos^2 \theta.
\end{eqnarray}
\end{subequations}
Here $m$ and $a$ correspond to the asymptotic mass and the spin parameter, respectively. The horizon of the Kerr black hole is a coordinate sphere $r = r_\mathrm{s}$, where
\begin{equation}
\label{rs}
r_\mathrm{s} \equiv \frac{1}{2}\sqrt{m^2 - a^2}.
\end{equation}
In the presence of the torus, the solution differs from the Kerr metric. Nevertheless, the Kerr solution still plays an important role in Shibata's formulation used in this work.

Following \cite{shibata}, we use the puncture formalism. Let $m$ and $a$ be parameters. We define $r_\mathrm{s}$ by Eq.\ (\ref{rs}), and replace the functions $\psi$ and $\alpha$ by $\phi$ and $B$ according to
\begin{equation}
\label{puncture}
\psi = \left( 1 + \frac{r_\mathrm{s}}{r} \right) e^\phi, \quad \alpha \psi = \left( 1 - \frac{r_\mathrm{s}}{r} \right) e^{-\phi} B.
\end{equation}
In the following, we will impose boundary conditions at $r = r_\mathrm{s}$ ensuring that this coordinate sphere is a minimal surface and an apparent horizon.

The shift vector is expressed as $\beta = \beta_\mathrm{K} + \beta_\mathrm{T}$, where $\beta_\mathrm{K}$ and $\beta_\mathrm{T}$ are defined as follows. We write the non-vanishing components of the extrinsic curvature tensor of hypersurfaces of constant time $t$ as
\begin{equation}
\label{Krf}
K_{r \varphi} = K_{\varphi r} = \frac{H_\mathrm{E} \sin^2 \theta}{\psi^2 r^2} +  \frac{1}{2 \alpha} \psi^4 r^2 \sin^2 \theta \partial_r \beta_\mathrm{T},
\end{equation}
\begin{equation}
\label{Ktf}
K_{\theta \varphi} = K_{\varphi \theta} = \frac{H_\mathrm{F} \sin \theta}{\psi^2 r} + \frac{1}{2 \alpha} \psi^4 r^2 \sin^2 \theta \partial_\theta \beta_\mathrm{T},
\end{equation}
where $H_\mathrm{E}$ and $H_\mathrm{F}$ are defined as
\begin{subequations}
\begin{eqnarray}
H_\mathrm{E} & = & \frac{ma \left[ (r_\mathrm{K}^2 - a^2) \Sigma_\mathrm{K} + 2 r_\mathrm{K}^2 (r_\mathrm{K}^2 + a^2) \right]}{\Sigma_\mathrm{K}^2}, \\
H_\mathrm{F} & = & - \frac{2 m a^3 r_\mathrm{K} \sqrt{\Delta_\mathrm{K}} \cos \theta \sin^2 \theta}{\Sigma_\mathrm{K}^2}.
\end{eqnarray}
\end{subequations}
Since for the Kerr spacetime $\beta_\mathrm{T} = 0$, one can view $\beta_\mathrm{K}$ as associated with the black hole and $\beta_\mathrm{T}$ as a contribution due to the torus.

We write the Einstein equations in the form of a set of equations for $q$, $\phi$, $B$ and $\beta_\mathrm{T}$:
\begin{widetext}
\begin{subequations}
\label{main_sys}
\begin{eqnarray}
\left[ \partial_{rr} + \frac{1}{r } \partial_r  + \frac{1}{r^2} \partial_{\theta \theta}  \right] q & = & S_q, \label{47}\\
\left[ \partial_{rr} + \frac{2 r  }{r^2 - r_\mathrm{s}^2} \partial_r + \frac{1}{r^2} \partial_{\theta \theta} + \frac{  \cot{\theta}}{r^2}  \partial_\theta \right] \phi & = & S_\phi, \label{44} \\
\left[ \partial_{rr} + \frac{3 r^2 +  r_\mathrm{s}^2}{r(r^2 - r_\mathrm{s}^2)} \partial_r + \frac{1}{r^2} \partial_{\theta \theta} + \frac{2 \cot{\theta}}{r^2}  \partial_\theta \right] B & = & S_B, \label{45} \\
\left[ \partial_{rr} + \frac{4 r^2 - 8 r_\mathrm{s} r + 2 r_\mathrm{s}^2}{r(r^2 - r_\mathrm{s}^2)} \partial_r + \frac{1}{r^2} \partial_{\theta \theta} + \frac{3 \cot{\theta}}{r^2}  \partial_\theta \right]  \beta_\mathrm{T} & = & S_{\beta_\mathrm{T}},  \label{46}
\end{eqnarray}
 \end{subequations}
where
\begin{subequations}
\label{sources}
\begin{eqnarray}
S_q & = & -8 \pi e^{2q} \left( \psi^4 p - \frac{\rho h u_\phi^2}{r^2 \sin^2 \theta} + \frac{3}{2} \psi^4 b^2 \right) + \frac{3 A^2}{\psi^8} + 2 \left[ \frac{r - r_\mathrm{s}}{r(r + r_\mathrm{s})} \partial_r + \frac{\cot \theta}{r^2} \partial_\theta \right] \tilde b \\
\nonumber & & + \left[ \frac{8 r_\mathrm{s}}{r^2 - r_\mathrm{s}^2} + 4 \partial_r (\tilde b - \phi) \right] \partial_r \phi + \frac{4}{r^2} \partial_\theta \phi \partial_\theta (\tilde b - \phi), \\
S_\phi & = & - 2 \pi e^{2q} \psi^4 \left[ \rho_\mathrm{H} - p + \frac{\rho h u_\phi^2}{\psi^4 r^2 \sin^2 \theta} - \frac{3}{2} b^2 \right] - \frac{A^2}{\psi^8} \\
\nonumber & & - \partial_r\phi \partial_r \tilde b - \frac{1}{r^2} \partial_\theta \phi \partial_\theta \tilde b - \frac{1}{2} \left[ \frac{r - r_\mathrm{s}}{r (r + r_\mathrm{s})} \partial_r \tilde b + \frac{\cot \theta}{r^2} \partial_\theta \tilde b \right], \\
S_B & = & 16 \pi B e^{2q} \psi^4 \left( p + \frac{1}{2} b^2 \right), \\
S_{\beta_\mathrm{T}} & = & \frac{16 \pi \alpha e^{2q} j_\varphi}{r^2 \sin^2 \theta} - 8 \partial_r \phi \partial_r \beta_\mathrm{T} + \partial_r \tilde b \partial_r \beta_\mathrm{T} - 8 \frac{\partial_\theta \phi \partial_\theta \beta_\mathrm{T}}{r^2} + \frac{\partial_\theta \tilde b \partial_\theta \beta_\mathrm{T}}{r^2}.
\end{eqnarray}
\end{subequations}
\end{widetext}
The shift component $\beta_\mathrm{K}$ satisfies the equation:
\begin{equation}
\label{betak_eq}
\partial_r \beta_\mathrm{K} = 2 H_\mathrm{E} B e^{-8 \phi} \frac{(r - r_\mathrm{s})r^2}{(r + r_\mathrm{s})^7}.
\end{equation}
Here, for convenience, we have introduced the variables $B = e^{\tilde b}$ and
\begin{equation}
\label{a2formula}
A^2 = \frac{(\psi^2 K_{r \varphi})^2}{r^2 \sin^2 \theta} + \frac{(\psi^2 K_{\theta \varphi})^2}{r^4 \sin^2 \theta},
\end{equation}
and defined
\begin{equation}
\rho_\mathrm{H} = \alpha^2 \rho h (u^t)^2 - p + \frac{1}{2} b^2,
\end{equation}
\begin{equation}
j_\varphi = \alpha \rho h u^t u_\varphi.
\end{equation}

The boundary conditions for the metric functions assumed at $r = r_\mathrm{s}$ are as follows:
\begin{equation}
\partial_r q = \partial_r \phi = \partial_r B = \partial_r \beta_\mathrm{T} = 0.
\end{equation}
It was noticed in \cite{shibata} that Eq.\ (\ref{46}) allows for a stronger condition. Following \cite{shibata} we require that $\beta_\mathrm{T} = O[(r - r_\mathrm{s})^4]$, or equivalently $\beta_\mathrm{T} = \partial_r \beta_\mathrm{T} = \partial_{rr} \beta_\mathrm{T} = \partial_{rrr} \beta_\mathrm{T} = 0$ at $r = r_\mathrm{s}$. Under the above boundary conditions, the two-surface $r = r_\mathrm{s}$ embedded in a hypersurface of constant time is a marginally outer trapped surface or the so-called apparent horizon.

\subsection{Keplerian rotation law and the prescription of the magnetic field}

Equation (\ref{bernoulli3}) requires a specification of two functions: $j(\Omega)$, the so-called rotation law, and
\begin{equation}
\label{magnetlaw}
b^2|\mathcal L| \equiv f(x), \quad x = \rho h |\mathcal L|,
\end{equation}
to which we refer as the magnetization law. 
 
We assume the Keplerian rotation law of the form
\begin{equation}
\label{keplerian_rl}
j(\Omega) = - \frac{1}{2} \frac{d}{d \Omega} \ln \left\{ 1 - \left[ a^2 \Omega^2 + 3 w^\frac{4}{3} \Omega^\frac{2}{3} (1 - a \Omega)^\frac{4}{3} \right] \right\},
\end{equation}
where $w$ is a free constant. This rotation law was derived in \cite{kkmmop, kkmmop2} and used subsequently in \cite{magnetic_nasze, OJ, kmm2019, ergosfery}. The main motivation behind Eq.\ (\ref{keplerian_rl}) is that this relation between $j$ and $\Omega$ is satisfied in the circular geodesic motion at the equatorial plane of the Kerr spacetime. That means, in particular, that a massless disk of dust around the Kerr black hole rotates according to Eq.\ (\ref{keplerian_rl}), in which case $w^2 = m$, where $m$ is the black hole mass. This rotation law proved to be a robust prescription also for self-gravitating fluids, but this time $w^2 \neq m$. In numerical computations, the value of $w$ is obtained from the requirement that the inner and outer coordinate radii of the disk read $R_1$ and $R_2$, respectively.

A Newtonian limit of Eq.\ (\ref{keplerian_rl}) yields the standard prescription for the angular velocity: $\Omega = w/(r \sin \theta)^\frac{3}{2}$. Newtonian models of self-gravitating disks obeying this standard Keplerian rotation law were investigated in \cite{mmp2013}.

The angular velocity within the disk can be computed from the relation $j(\Omega) = u^t u_\varphi$, which in more explicit terms reads
\begin{equation}
\label{rot_law_eq}
j(\Omega) \left[ \alpha^2 - \psi^4 r^2 \sin^2 \theta (\Omega + \beta)^2 \right] = \psi^4 r^2 \sin^2 \theta (\Omega + \beta).
\end{equation}

In prescribing the magnetization law (\ref{magnetlaw}) we follow \cite{magnetic_nasze} and choose
\begin{equation}
\label{magnetizationlaw}
f(x) = 2 n \left[ x - \frac{1}{C_1} \ln (1 + C_1 x) \right],
\end{equation}
where $n$ and $C_1$ are constant parameters. This yields
\begin{equation} 
\int \frac{d (b^2 \mathcal L)}{2 \rho h \mathcal L} = \ln \left[ \left( 1 + C_1 \alpha^2 \psi^4 r^2 \sin^2 \theta \rho h \right)^n \right] . 
\end{equation}
Note that $b^2 = 0$ for $\rho = 0$.

Assuming the above choices, one can write the Euler-Bernoulli Eq.\ (\ref{bernoulli3}) as
\begin{eqnarray}
\label{bernoulli2}
\lefteqn{h \left( 1 + C_1 \alpha^2 \psi^4 r^2 \sin^2 \theta \rho h \right)^n } \nonumber \\
&& \times \sqrt{ \alpha^2 - \psi^4 r^2 \sin^2 \theta (\Omega + \beta)^2} \nonumber \\
&& \times \left\{ 1 - \left[ a^2 \Omega^2 + 3 w^\frac{4}{3} \Omega^\frac{2}{3} (1 - a \Omega)^\frac{4}{3} \right] \right\}^{-\frac{1}{2}} = C^\prime.
\end{eqnarray}

We assume a polytropic equation of state $p = K\rho^\Gamma$, where $K$ and $\Gamma$ are constant. The specific enthalpy is then given by
\begin{equation}
\label{polytropic_h}
h = 1 + \frac{K \Gamma}{\Gamma - 1} \rho^{\Gamma - 1}.
\end{equation}

\subsection{Masses and angular momenta}

The most natural mass measure of the black hole-torus system is the total Arnowitt-Deser-Misner (ADM) asymptotic mass. For numerical convenience, we compute the ADM mass as
\begin{equation} 
m_\mathrm{ADM} = \sqrt{m^2 - a^2} + M_1, 
\end{equation}
where
\begin{equation}
\label{m1}
M_1 = - 2 \int_{r_\mathrm{s}}^\infty dr \int_0^{\pi/2} d \theta (r^2 - r_\mathrm{s}^2) \sin \theta S_\phi,
\end{equation}
i.e., as a volume integral instead of an asymptotic one.

The black hole mass can be defined in many ways. Following \cite{shibata}, we use Christodoulou's formula \cite{christodoulou}
\begin{equation}
 M_\mathrm{BH} = M_\mathrm{irr} \sqrt{1 + \frac{J_\mathrm{H}^2}{4 M_\mathrm{irr}^4}}.
\end{equation}
Here $J_\mathrm{H}$ is the angular momentum of the black hole given by
\begin{equation}
\label{jhdef}
J_\mathrm{H} = \frac{1}{4} \int_0^{\pi/2} d \theta \left( \frac{r^4 \sin^3 \theta \psi^6 \partial_r \beta}{\alpha} \right)_{r=r_\mathrm{s}}, 
\end{equation}
and $M_\mathrm{irr}$ is the so-called irreducible mass:
\begin{equation}
M_\mathrm{irr} = \sqrt{\frac{A_\mathrm{H}}{16 \pi}},
\end{equation}
where $A_\mathrm{H}$ is the area of the horizon:
\begin{equation}
A_\mathrm{H} = 4 \pi \int_0^{\pi/2} d \theta \left( \psi^4 e^q r^2 \sin \theta \right)_{r=r_\mathrm{s}}.
\end{equation}

The angular momentum of the torus is defined in a standard way following from the conservation law $\eta^\nu \nabla_\mu T\indices{^\mu_\nu} = \nabla_\mu (T\indices{^\mu_\nu} \eta^\nu) = 0$, where $\eta^\mu = (0,0,0,1)$ is the axial Killing vector \cite{leshouches}. This yields the expression for the angular momentum of the torus in the form
\begin{eqnarray}
J_1 & = & \int \sqrt{- g} T\indices{^t_\varphi} d^3 x \nonumber \\
& = & 4 \pi \int_{r_\mathrm{s}}^\infty dr \int_0^{\pi/2} d\theta r^2 \sin \theta \alpha \psi^6 e^{2q} \rho h u^t u_\varphi.
\label{j1}
\end{eqnarray}
The total angular momentum reads
\begin{equation}
J = J_\mathrm{H} + J_1.
\end{equation}

The angular momentum defined by Eq.\ (\ref{jhdef}) depends on the boundary conditions assumed at the black hole horizon. In our case, the conditions $\beta_\mathrm{T} = \partial_r \beta_\mathrm{T} = \partial_{rr} \beta_\mathrm{T} = \partial_{rrr} \beta_\mathrm{T} = 0$ at $r = r_\mathrm{s}$ yield $J_\mathrm{H} = a m$. 

\subsection{Parametrization of solutions and the numerical scheme}

There are a few ways of parametrizing the solutions. Here we follow the choice of \cite{kkmmop,kkmmop2,magnetic_nasze}. We specify the black hole parameters $m$ and $a$, the inner and outer equatorial coordinate radii of the torus $R_1$ and $R_2$, the polytropic exponent $\Gamma$, the maximal rest-mass density within the torus $\rho_\mathrm{max}$, and, in the case of magnetized tori, parameters $n$ and $C_1$.

In the remainder of this paper we set $m = 1$. That means that the parameter $m$ can be also treated as a unit of mass and length. In practice, if the torus is sufficiently light, the mass of the black hole $M_\mathrm{BH}$ is well approximated by $m$.

The numerical scheme used in this paper to solve Eqs.\ (\ref{main_sys}), (\ref{betak_eq}), (\ref{rot_law_eq}), and (\ref{bernoulli2}) is essentially the same as in Ref.\ \cite{magnetic_nasze}. It is an iterative, finite difference code. An improvement with respect to the version used in \cite{magnetic_nasze} consists of using the PARDISO linear algebra library \cite{pardiso}, instead of LAPACK \cite{lapack}, which was used previously.

The numerical grid spans over a finite spatial region $r_\mathrm{s} \le r \le r_\infty$, $0 \le \theta \le \pi/2$, where $r_\infty$ is large, but finite. The boundary conditions assumed at $r = r_\infty$ stem from the asymptotic expansion of the metric functions. In particular, we assume at $r = r_\infty$:
\begin{subequations}
\begin{eqnarray}
    \phi & \sim & \frac{M_1}{2r},\\
    \beta_\mathrm{T} & \sim & - \frac{2 J_1}{r^3},\\
    B & \sim & 1 - \frac{B_1}{r^2}, \\
    q & \sim & \frac{q_1 \sin^2 \theta}{r^2},
\end{eqnarray}
\end{subequations}
where $M_1$ and $J_1$ are given by Eqs.\ (\ref{m1}) and (\ref{j1}), and $B_1$ and $q_1$ are computed as
\begin{equation}
    B_1  = \frac{2}{\pi} \int_{r_\mathrm{s}}^\infty dr \frac{(r^2 - r_\mathrm{s}^2)^2}{r} \int_0^{\pi/2} d \theta \sin^2 \theta S_B
\end{equation}
and
\begin{eqnarray}
    q_1 & = & \frac{2}{\pi} \int_{r_\mathrm{s}}^\infty dr r^3 \int_0^{\pi/2} d\theta \cos(2 \theta) S_q \nonumber \\
    && - \frac{4 r_\mathrm{s}^2}{\pi} \int_0^{\pi/2} d \theta \cos(2 \theta) q(r_\mathrm{s}, \theta).
\end{eqnarray}

The remainder of this paper is devoted to the discussion of numerical solutions. Since we focus mainly on the impact of the magnetic field on black hole-torus configurations, we set in all our examples $\Gamma = 4/3$. Moreover, we only consider co-rotating tori ($a \ge 0$, $\Omega > 0$), as they seem to be much more realistic astrophysically.

\section{Configurations with low-mass tori}
\label{lowmass}

\begin{figure}
\includegraphics[width=\columnwidth]{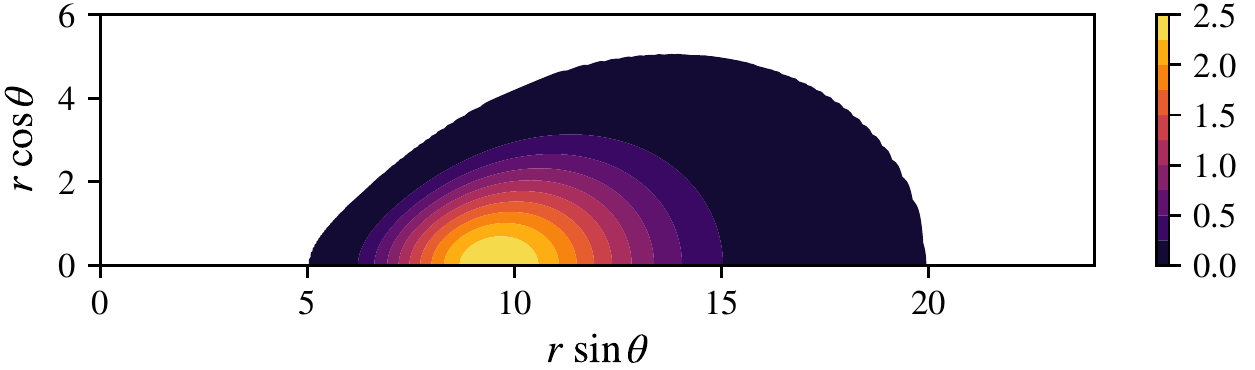}
\\ \vspace{-1.8em}
\includegraphics[width=\columnwidth]{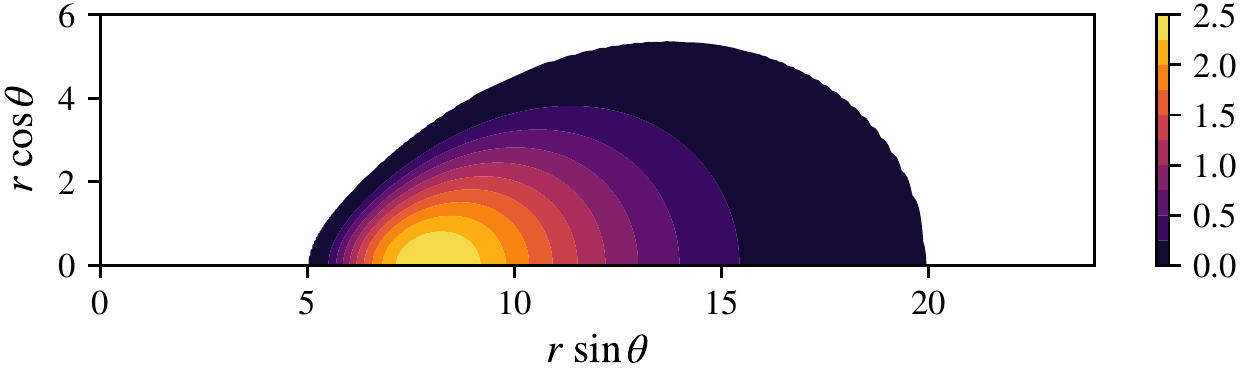}
\\ \vspace{-1.8em}
\includegraphics[width=\columnwidth]{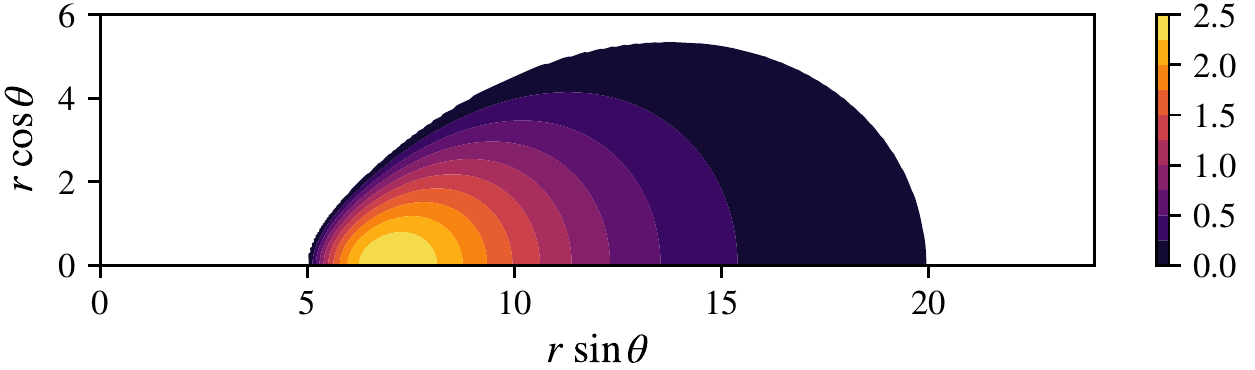}
    \caption{\label{fig:profil_n1.5a.9p2.5} Distribution of the rescaled baryonic mass density $10^4\rho$ for
    parameter values $n=1.5$, $a=0.9$, $\rho_{\rm max}=2.5 \times 10^{-4}$ and successive values of the magnetic field parameter $C_1=0.1$ (top), $C_1=0.5$ (center), and $C_1=1.0$ (bottom).}
\end{figure}

\begin{figure}
\includegraphics[width=\columnwidth]{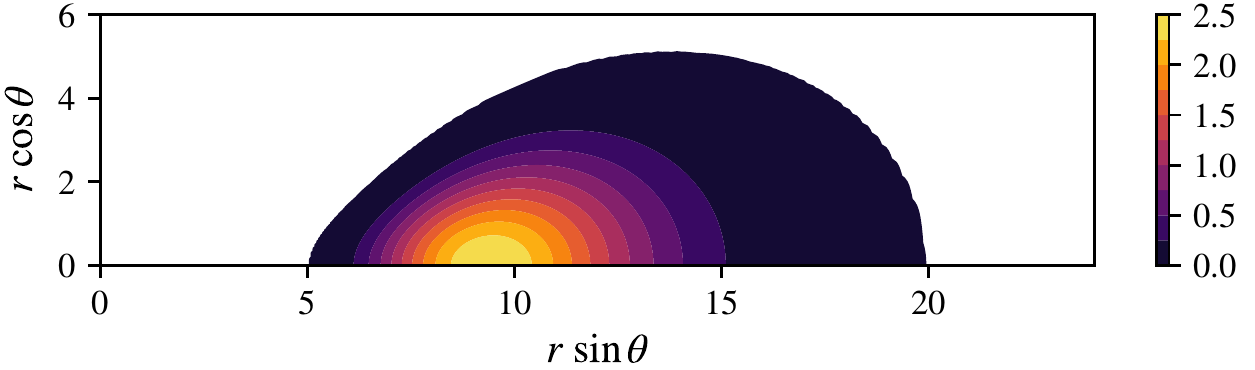}
    \\ \vspace{-1.8em}
\includegraphics[width=\columnwidth]{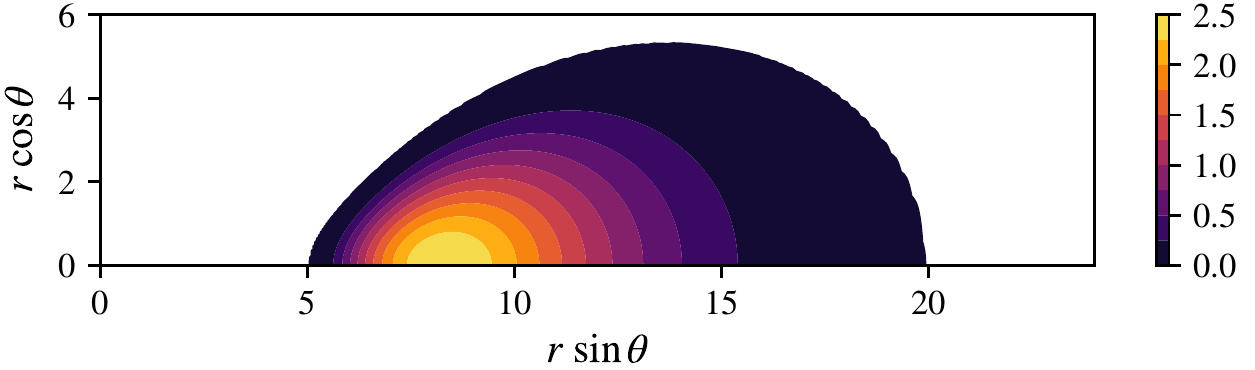}
    \\ \vspace{-1.8em}
\includegraphics[width=\columnwidth]{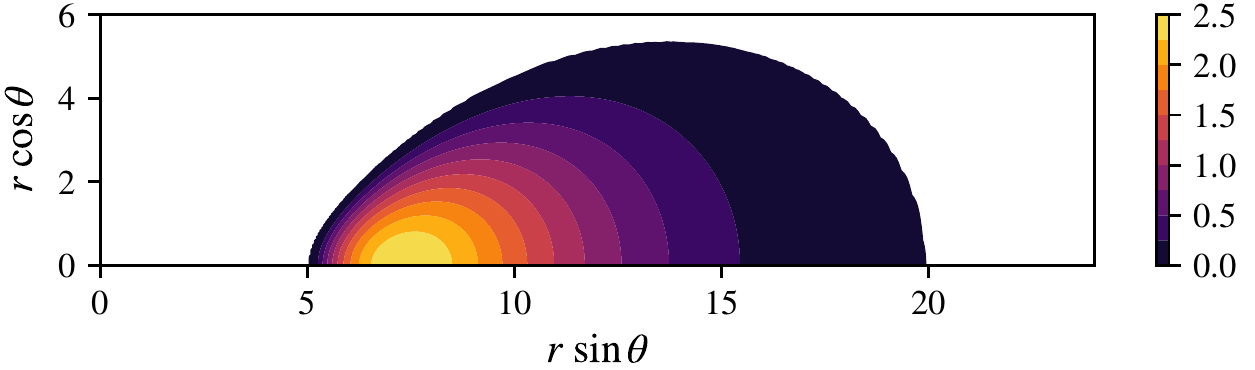}
    \caption{\label{fig:profil_n2a.9p2.5} Distribution of the rescaled baryonic mass density $10^4\rho$ for
    parameter values $n = 2$, $a = 0.9$, $\rho_{\rm max}=2.5 \times 10^{-4}$ and successive values of the magnetic field parameter $C_1=0.1$ (top), $C_1=0.3$ (center), and $C_1=0.6$ (bottom).}
\end{figure}

\begin{figure}
\includegraphics[width=\columnwidth]{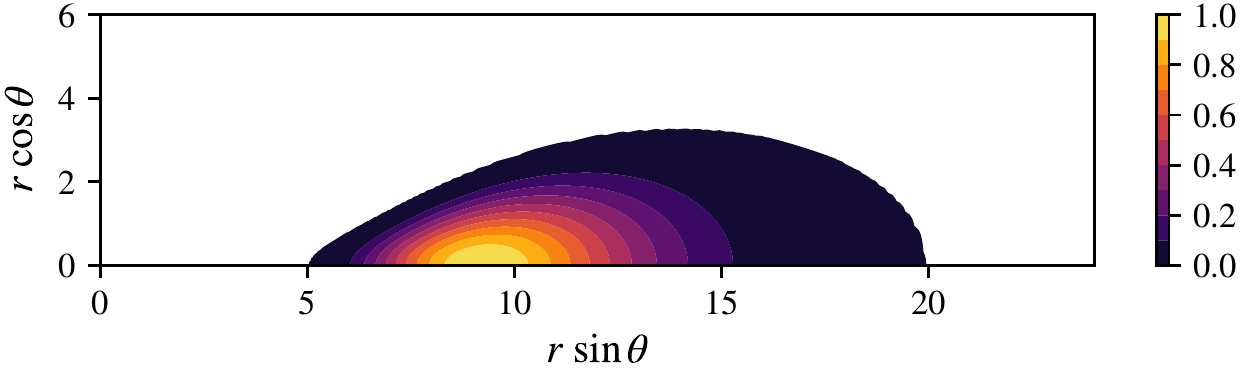}
    \\ \vspace{-1.8em}
\includegraphics[width=\columnwidth]{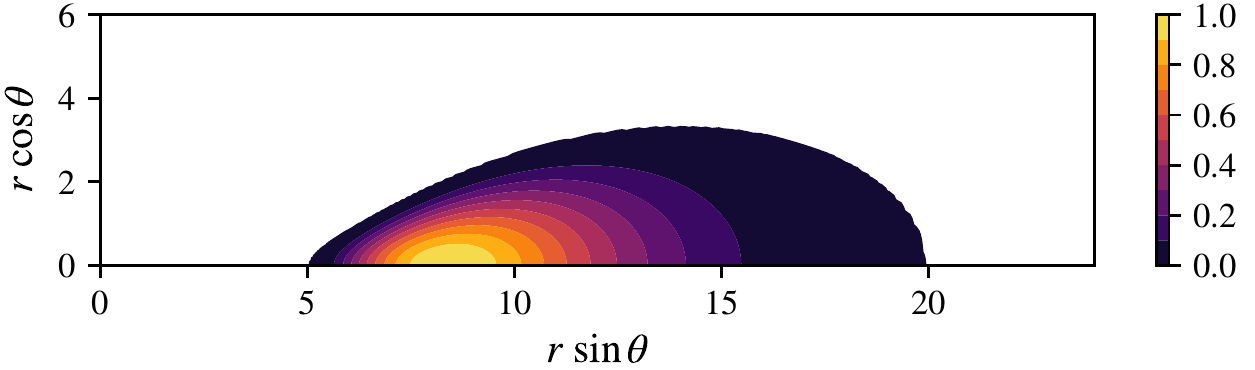}
    \\ \vspace{-1.8em}
\includegraphics[width=\columnwidth]{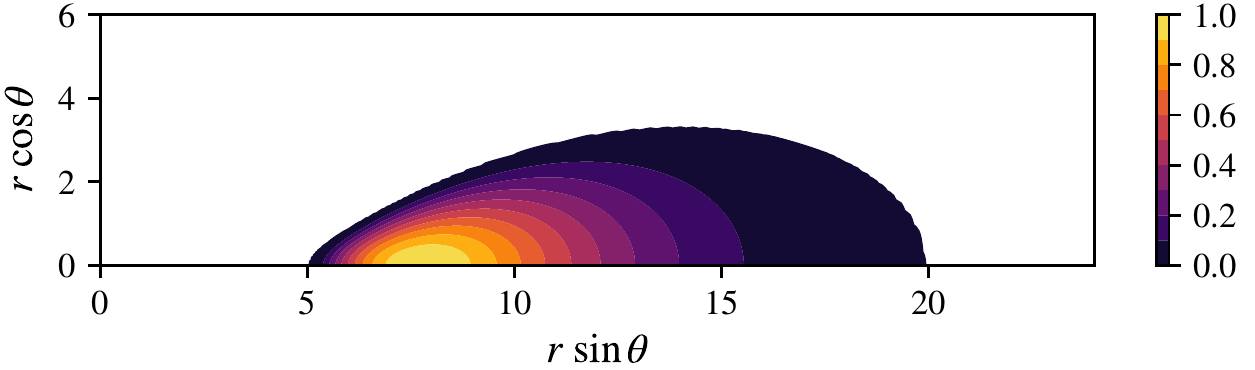}
    \\ \vspace{-1.8em}
\includegraphics[width=\columnwidth]{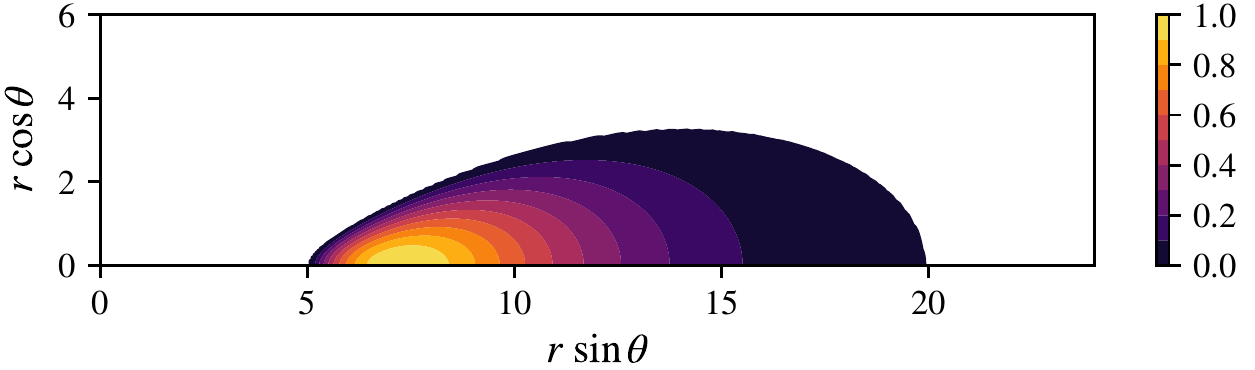}
    \caption{ \label{fig:profil_n2a.9p1} Distribution of the rescaled baryonic mass density $10^4\rho$ for
    parameter values $n=2$, $a=0.9$, $\rho_\mathrm{max}=10^{-4}$
    and successive values of the magnetic field parameter $C_1=0.1$ (top), $C_1=0.2$ (higher center), $C_1=0.3$ (lower center), and $C_1=0.4$ (bottom).}
\end{figure}

\begin{figure}
\includegraphics[width=\columnwidth]{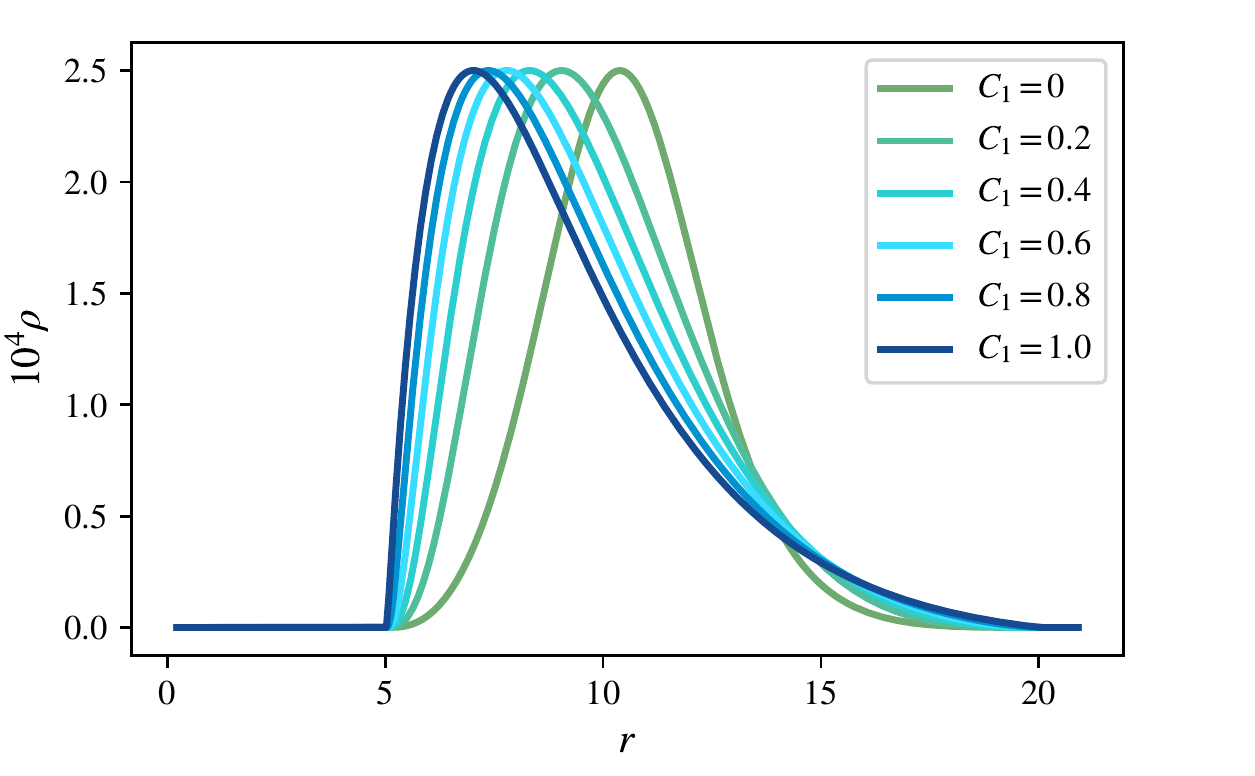}
    \caption{\label{fig:rho_n1.5a.9p2.5} Distribution of the rescaled baryonic mass density $10^4\rho$ in the 
    equatorial plane $\theta=\pi/2$, for successive values of the parameter $C_1$ and fixed $n=1.5$, $a=0.9$, and $\rho_\mathrm{max}=2.5 \times 10^{-4}$.}
\end{figure}

\begin{figure}
\includegraphics[width=\columnwidth]{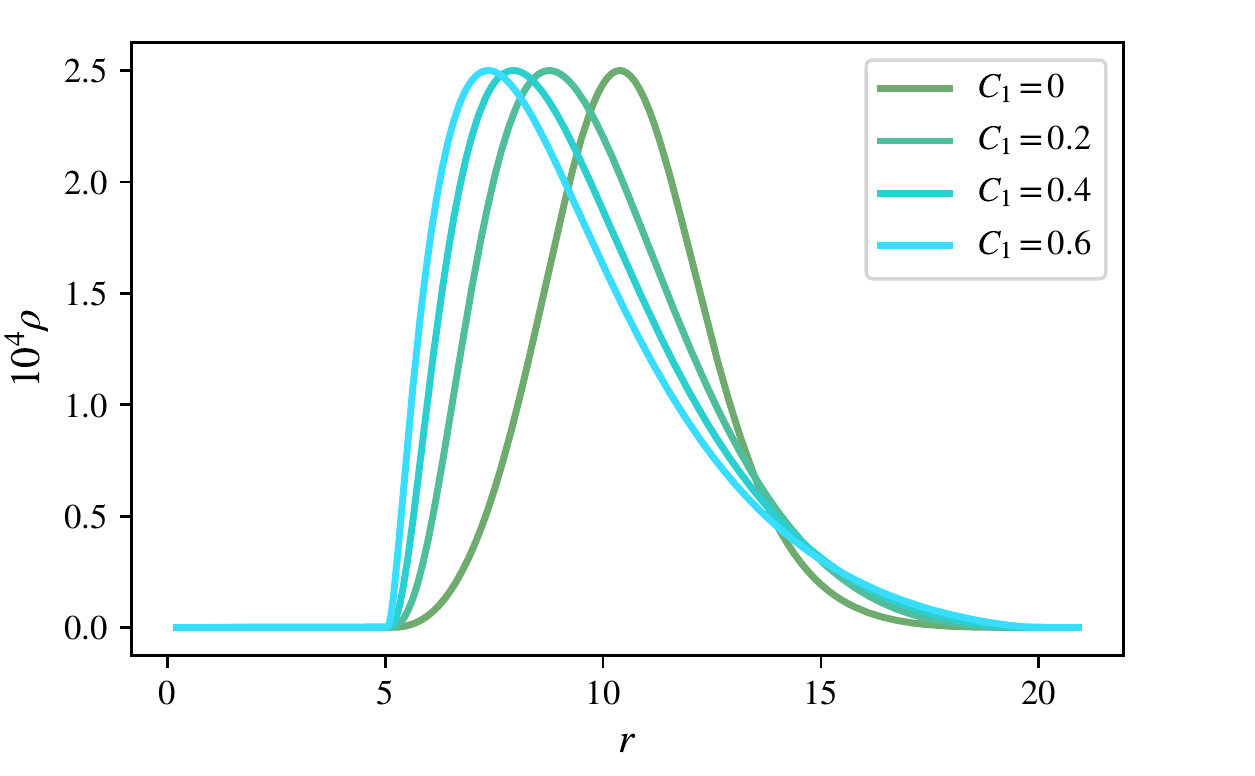}
    \caption{\label{fig:rho_n2a.9p2.5} Distribution of the rescaled baryonic mass density $10^4\rho$ in the 
    equatorial plane $\theta=\pi/2$, for successive values of the parameter $C_1$ and fixed $n=2$, $a=0.9$, and $\rho_\mathrm{max}=2.5 \times 10^{-4}$.}
\end{figure}

\begin{figure}
\includegraphics[width=\columnwidth]{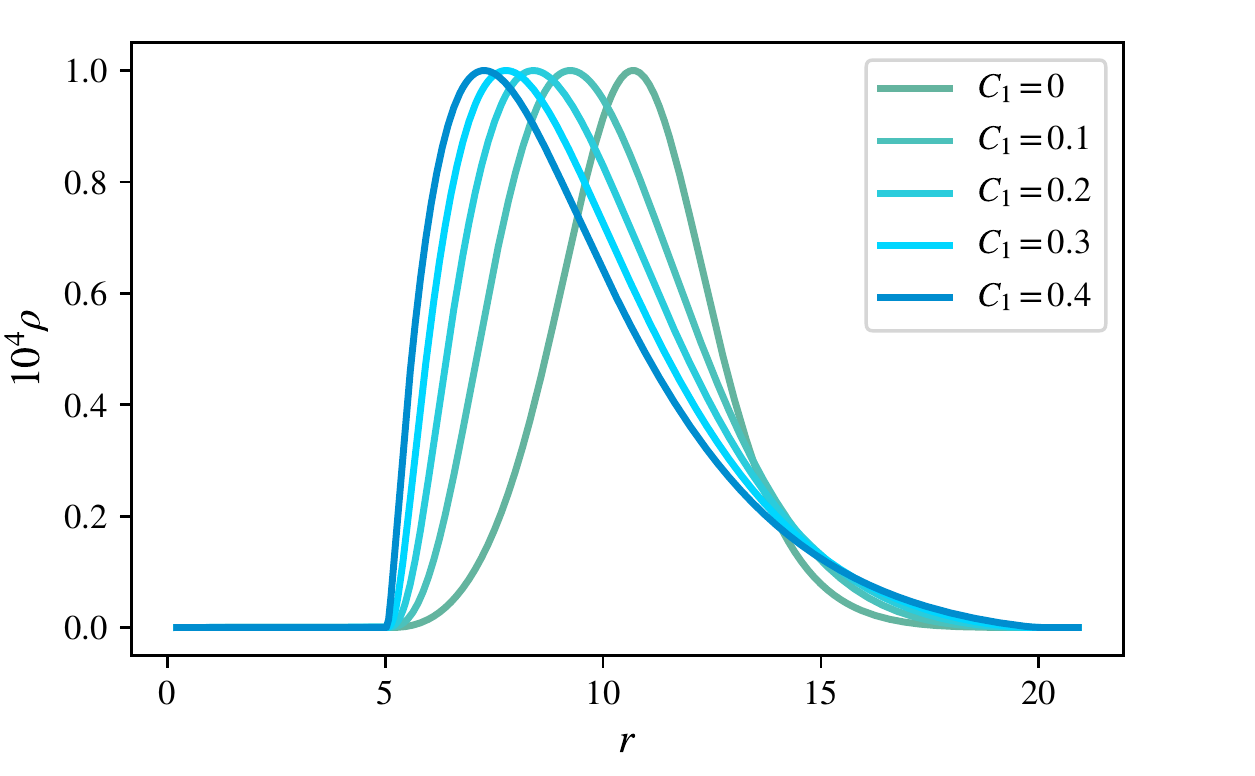}
    \caption{\label{fig:rho_n2a.9p1} Distribution of the rescaled baryonic mass density $10^4\rho$ in the 
    equatorial plane $\theta=\pi/2$, for successive values of the parameter $C_1$ and fixed $n=2$, $a=0.9$, and $\rho_\mathrm{max}=10^{-4}$.}
\end{figure}

\begin{figure}
\includegraphics[width=\columnwidth]{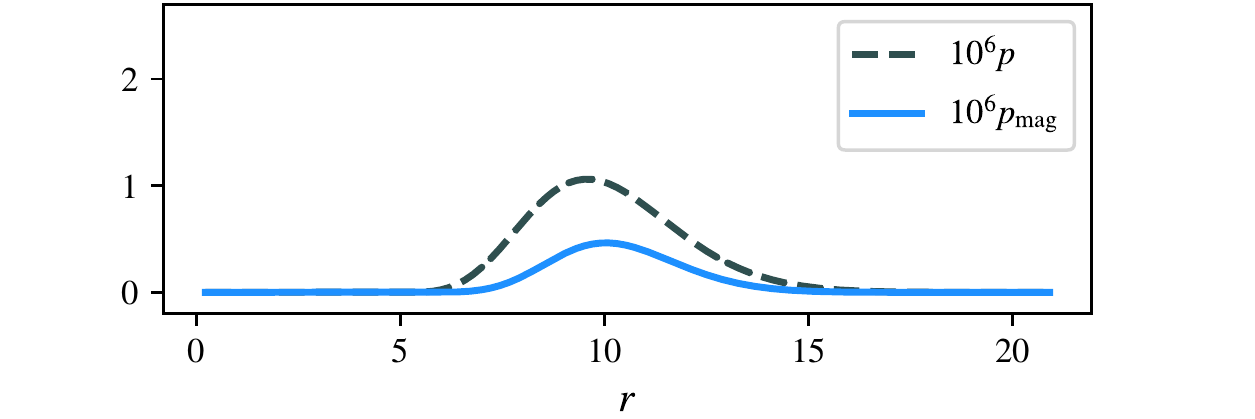}
    \\ \vspace{-2em}
\includegraphics[width=\columnwidth]{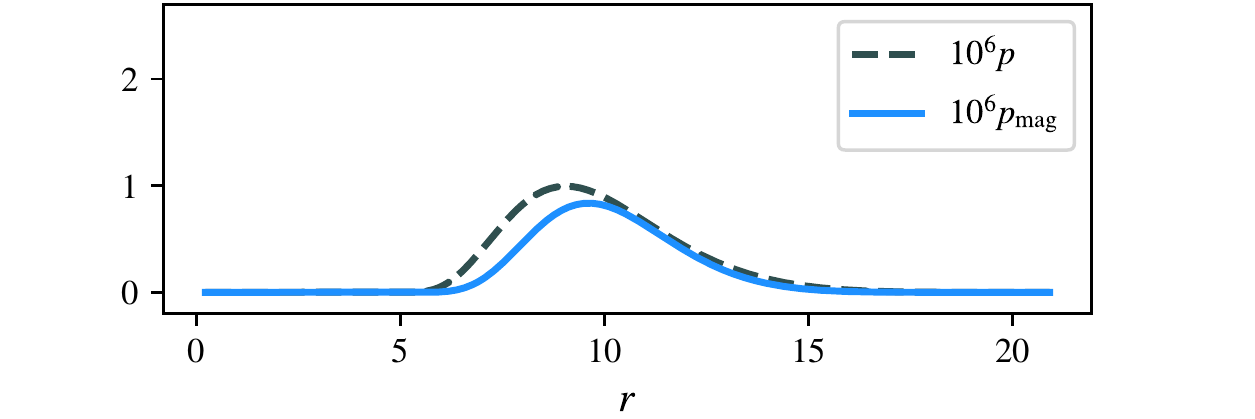}
    \\ \vspace{-2em}
\includegraphics[width=\columnwidth]{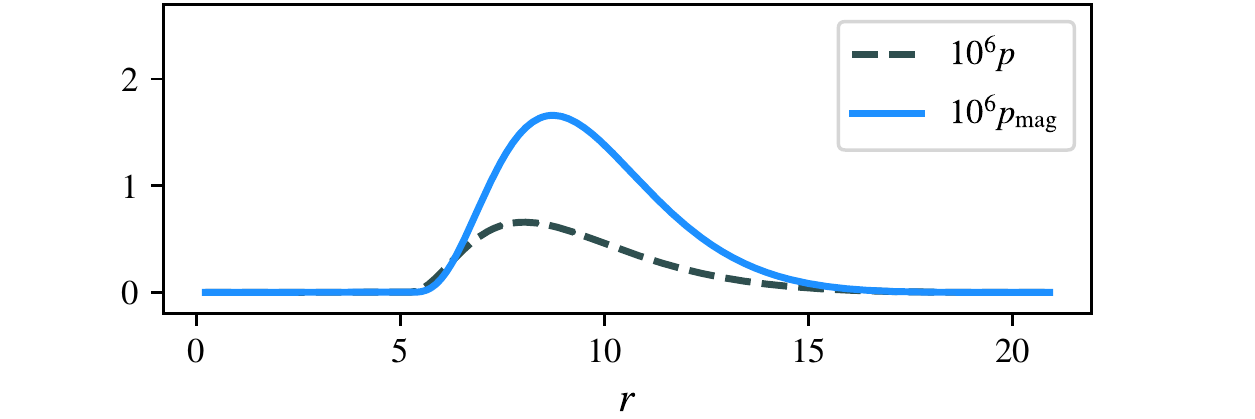}
    \\ \vspace{-2em}
\includegraphics[width=\columnwidth]{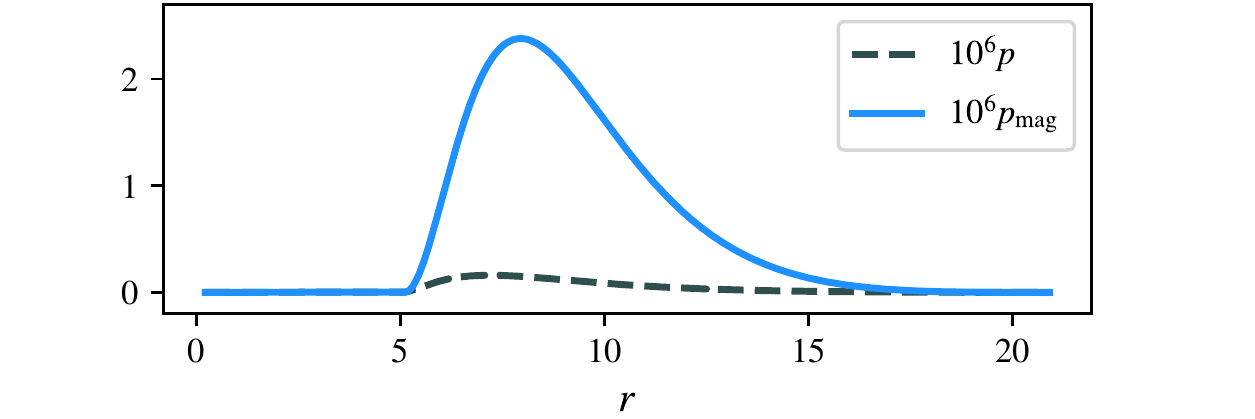}
    \caption{\label{fig:pressure_n1.5a.9p2.5} Thermal pressure $p$ and the magnetic pressure $p_\mathrm{mag}$ in the equatorial plane. Here $n = 1.5$, $a = 0.9$, and $\rho_\mathrm{max}=2.5 \times 10^{-4}$. Upper panel: $C_1 = 0.1$; upper central panel: 
    $C_1 = 0.2$; lower central panel: $C_1 = 0.5$; lower panel: $C_1 = 0.9$.}  
\end{figure}

\begin{figure}
\includegraphics[width=\columnwidth]{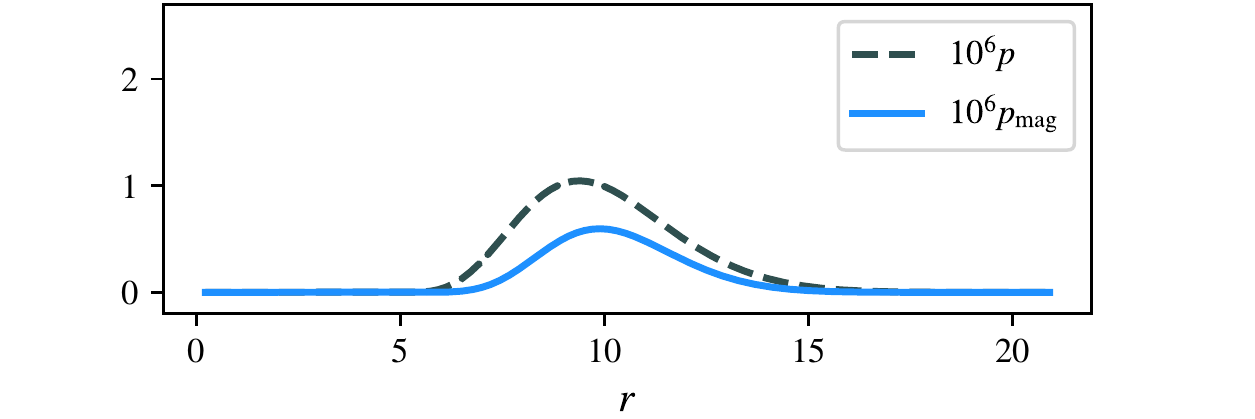}
    \\ \vspace{-2em}
\includegraphics[width=\columnwidth]{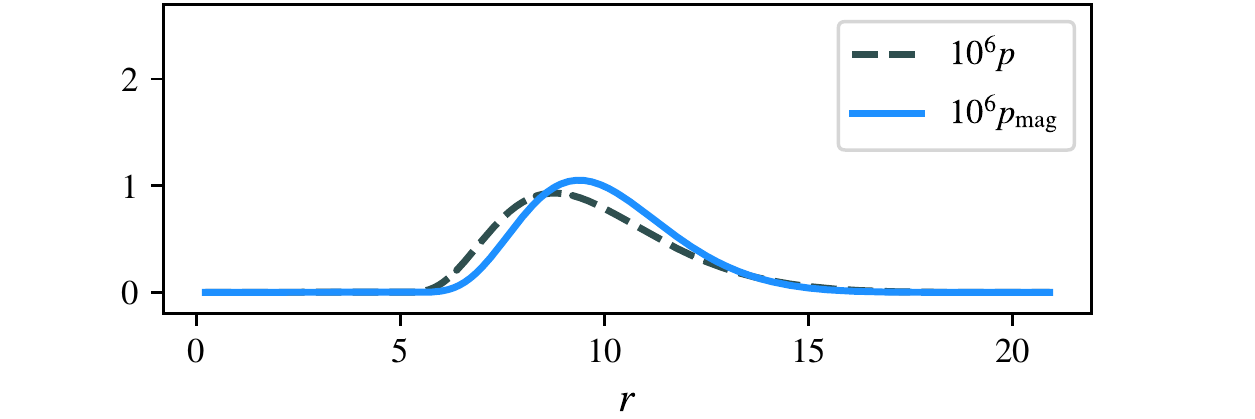}
    \\ \vspace{-2em}
\includegraphics[width=\columnwidth]{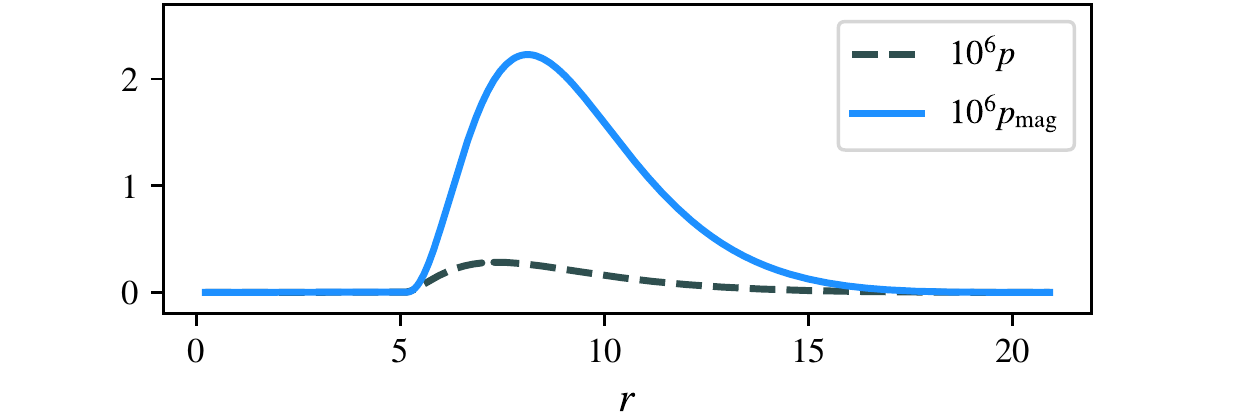}
    \caption{\label{fig:pressure_n2a.9p2.5} Thermal pressure $p$ and the magnetic pressure $p_\mathrm{mag}$ in the equatorial plane. Here $n = 2$, $a = 0.9$, and $\rho_\mathrm{max}=2.5 \times 10^{-4}$. Upper panel: $C_1 = 0.1$; central panel: 
    $C_1 = 0.2$; lower panel: $C_1 = 0.6$.}
\end{figure}

\begin{figure}
\includegraphics[width=\columnwidth]{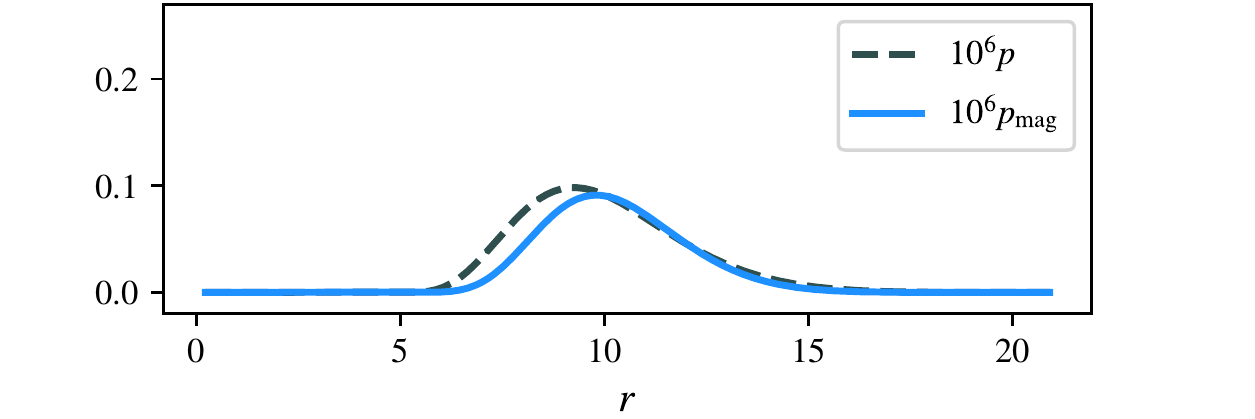}
    \\ \vspace{-2em}
\includegraphics[width=\columnwidth]{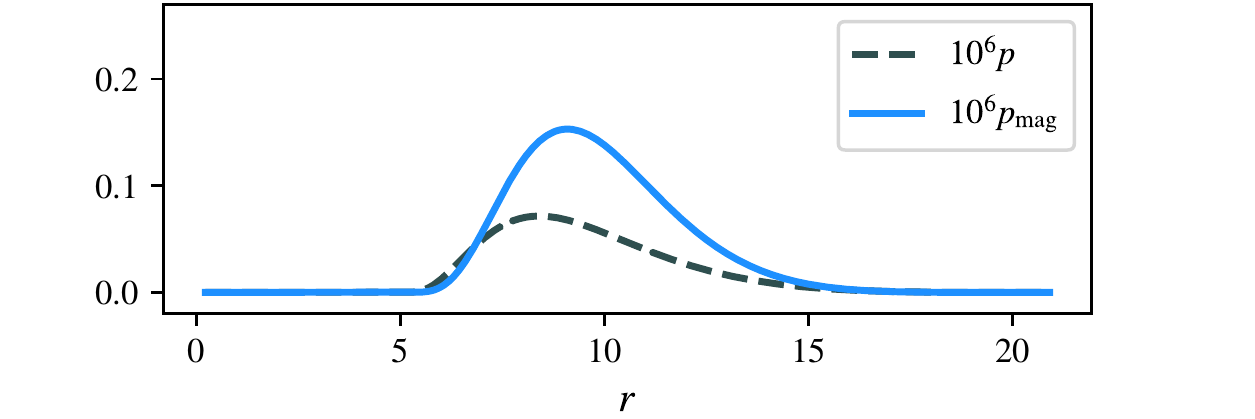}
    \\ \vspace{-2em}
\includegraphics[width=\columnwidth]{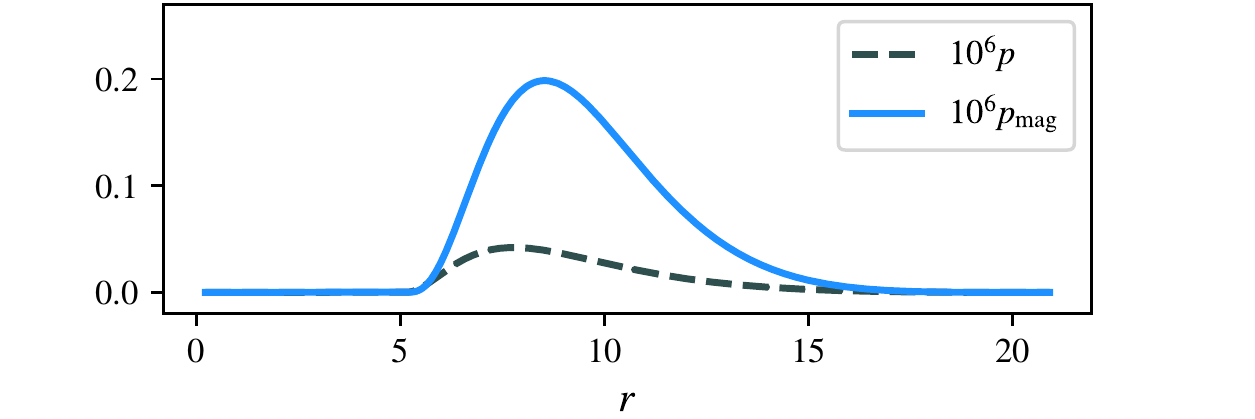}
    \caption{\label{fig:pressure_n2a.9p1} Thermal pressure $p$ and the magnetic pressure $p_\mathrm{mag}$ in the equatorial plane. Here $n = 2$, $a = 0.9$, and $\rho_\mathrm{max} = 10^{-4}$. Upper panel: $C_1 = 0.1$; central panel: 
    $C_1 = 0.2$; lower panel: $C_1 = 0.3$.}
\end{figure}

\begin{figure}
\includegraphics[width=\columnwidth]{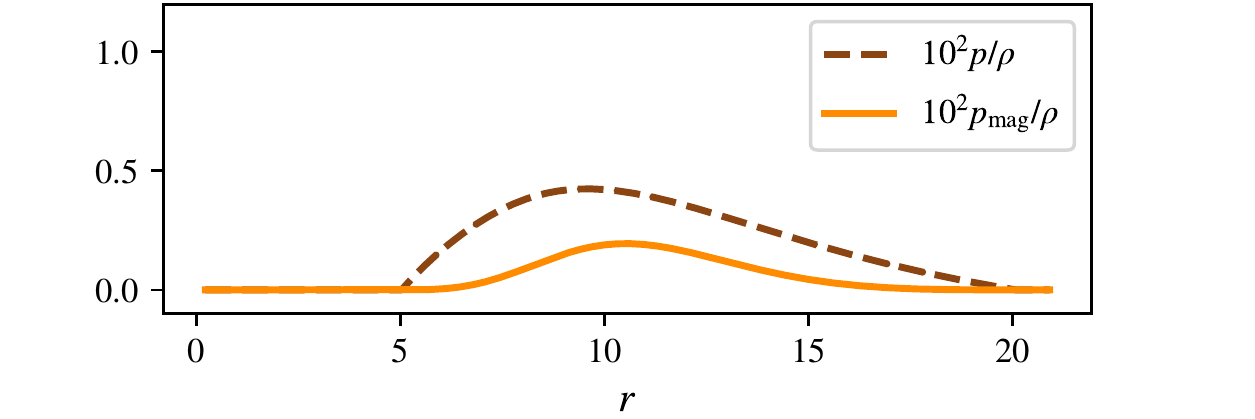}
    \\ \vspace{-2em}
\includegraphics[width=\columnwidth]{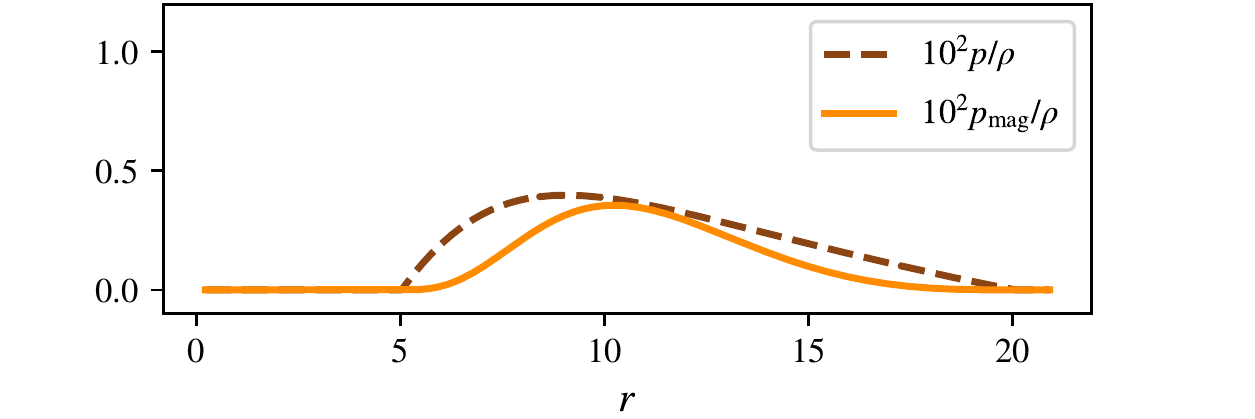}
    \\ \vspace{-2em}
\includegraphics[width=\columnwidth]{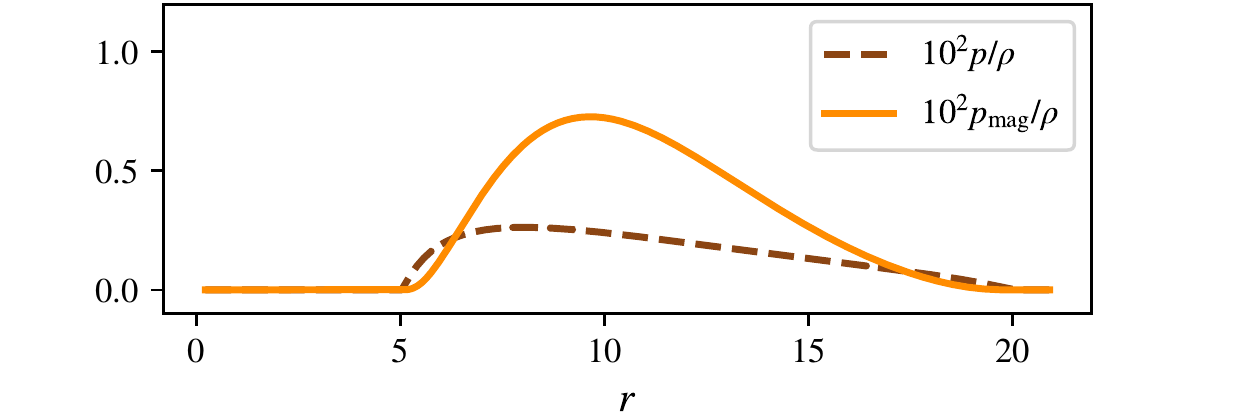}
    \\ \vspace{-2em}
\includegraphics[width=\columnwidth]{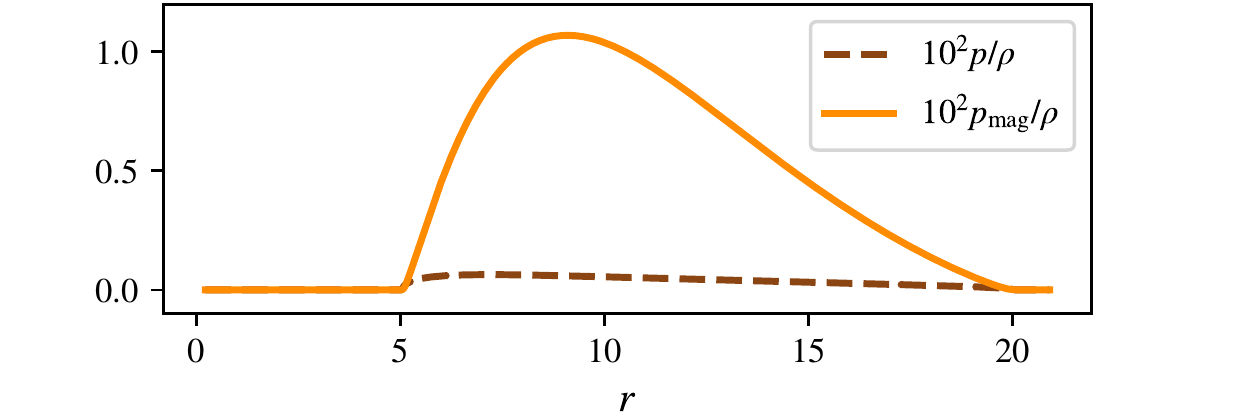}

\caption{\label{fig:pressureandrho} The ratios of $10^2 p/\rho$ and $10 ^2 p_\mathrm{mag}/\rho$ at the equatorial plane. The data correspond to solutions shown in Fig.\ \ref{fig:pressure_n1.5a.9p2.5}. Here $n = 1.5$, $a = 0.9$, and $\rho_\mathrm{max}=2.5 \times 10^{-4}$. Upper panel: $C_1 = 0.1$; upper central panel: $C_1 = 0.2$; lower central panel: $C_1 = 0.5$; lower panel: $C_1 = 0.9$.}
\end{figure}

We start our discussion with examples of relatively low-mass tori. Sample data describing such configurations are collected in Tables \ref{tab1}--\ref{tab4}. In these examples we set the black hole spin parameter $a = 0.5$ or $a = 0.9$ and inner and outer coordinate radii of the torus $(R_1, R_2) = (5,20)$, $(R_1, R_2) = (3,20)$,  or $(R_1, R_2) = (1.5,20)$. We focus on varying the parameters $n$ and $C_1$ in the magnetization law (\ref{magnetizationlaw}) and the maximum value of the rest-mass density in the disk $\rho_\mathrm{max}$. In principle, the choice of the parameter $n$ should be important, since it controls the assumed magnetization law. All examples given in Ref.\ \cite{magnetic_nasze} were computed assuming $n = 1$.

The tables report the mass of the black hole $M_\mathrm{BH}$, the total asymptotic mass of the system $m_\mathrm{ADM}$, the magnetization parameter $\beta_\mathrm{mag}$, the inner and outer circumferential radii of the torus $R_\mathrm{c1}$, $R_\mathrm{c2}$, the coordinate and circumferential radii of the Innermost Stable Circular Orbit (ISCO), denoted as $r_\mathrm{ISCO}$ and $r_\mathrm{c,ISCO}$, respectively.

In practice, the magnetic field component is controlled by two parameters, $n$ and $C_1$, but the resulting strength of the magnetic field depends on other parameters of the model. The magnetization parameter $\beta_\mathrm{mag}$ is defined as a ratio
\[ \beta_\mathrm{mag} = \frac{p}{p_\mathrm{mag}} = \frac{2 p}{b^2} \]
taken at the point in which the rest-mass density of the gas (or the thermal pressure $p$) attains its maximum. Thus $\beta_\mathrm{mag} = \infty$ corresponds to a non-magnetized fluid.

Circumferential radii are also defined in a standard way---a circle with a circumference $L$ has a circumferential radius $r_\mathrm{c} = L/(2\pi)$. In our case, the circumferential radius $r_\mathrm{c}$ and the coordinate radius $r$ of a circle of constant $t$, $r$, and $\theta$ are related by $r_\mathrm{c} = \psi^2 r \sin \theta$.

The ISCO is computed as described in \cite{ergosfery}. For massive disks the algebraic condition for the ISCO can be satisfied in multiple discrete locations. Consequently, we take the name ISCO literally, and only consider innermost orbits satisfying this condition.

Sample meridional profiles of the rest-mass density $\rho$ are shown in Figs.\ \ref{fig:profil_n1.5a.9p2.5}--\ref{fig:profil_n2a.9p1}. One of general effects, which can be observed in these figures and which was noted already in \cite{magnetic_nasze}, is that the location of the maximum of the rest-mass density within the disk is shifted toward smaller radii (i.e., toward the black-hole) for the increasing magnetic field. To make this effect more visible, we plot in Figs.\ \ref{fig:rho_n1.5a.9p2.5}--\ref{fig:rho_n2a.9p1} the rest-mass density $\rho$ at the equatorial plane for a selection of solutions from Table \ref{tab1}. A very general property illustrated in Figs.\ \ref{fig:profil_n1.5a.9p2.5}--\ref{fig:profil_n2a.9p1}, characteristic also for not magnetized Keplerian disks, is that their geometric thickness decreases with a decreasing mass of the disk. Very massive tori have roughly circular meridional cross sections (cf.\ Figs.\ \ref{profrhoErgbp}--\ref{profrhoErgbp3}).

For disks with sufficiently large inner radius $R_1$, the asymptotic mass $m_\mathrm{ADM}$ seems to grow with the increasing magnetic-field component. This is the case for the majority of models collected in Table \ref{tab1}, and also for the majority of massive disks discussed in the next section. The situation can be reversed for relatively light disks with small inner radii $R_1$. In Table \ref{tab3} we collect a small sample of solutions illustrated in Figs.\ \ref{bif2} and \ref{bif2bmag}. We group solutions characterized by the same asymptotic mass $m_\mathrm{ADM}$, but different $\rho_\mathrm{max}$ and different magnetization level. It turns out that $\rho_\mathrm{max}$ does not have to change monotonically with $\beta_\mathrm{mag}$.

Another morphological feature, observed already in \cite{magnetic_nasze} and confirmed in our present study, is the proportion between the magnetic pressure $p_\mathrm{mag} = \frac{1}{2}b^2$ and the thermal pressure $p$ among configurations differing in the magnetization, i.e., parameters $n$ and $C_1$. In general, by increasing the strength of the magnetic field we shift from configurations whose equilibrium structure is governed by the distribution of the thermal pressure to the ones where the dynamical role of counterbalancing the gravitational and centrifugal forces is played almost exclusively by the gradients of the magnetic pressure. On the other hand, in comparison to the rest-mass density, both magnetic and thermal pressures remain relatively small, with the ratios $p/\rho$ or $p_\mathrm{mag}/\rho$ of the order of $10^{-2}$ at most. Sample plots of the thermal and magnetic pressures at the equatorial plane are shown in Figs.\ \ref{fig:pressure_n1.5a.9p2.5}--\ref{fig:pressure_n2a.9p1}. For clarity, we decided to select for these plots the same families of models as in Figs.\ \ref{fig:profil_n1.5a.9p2.5}--\ref{fig:profil_n2a.9p1}. In Figure \ref{fig:pressureandrho} we plot the ratios $p/\rho$ and $p_\mathrm{mag}/\rho$ for solutions depicted in Fig.\ \ref{fig:pressure_n1.5a.9p2.5}.

Table \ref{tab2} requires a separate comment. Here, assuming $C_1 = 1$ and four sample values of the exponent $n = 0.5, 1, 1.5, 2$, we construct series of models with a decreasing asymptotic mass $m_\mathrm{ADM}$. There is a limit on the minimal mass of the torus, depending on the magnetic field contribution, below which we do not find numerical solutions. This is consistent with a common experience in modeling of rotating disks around black holes, where the existence of numerical solutions depends strongly on the assumed rotation law. Let us recall that no rigidly rotating compact disks are allowed in the Kerr spacetime \cite{AP}. The Keplerian rotation law (\ref{keplerian_rl}) allows for light compact disks, although they get geometrically thinner and thinner with a decreasing mass. It should be stressed that rotation law (\ref{keplerian_rl}) can be quite sensitive to a change of details. A limit of Eq.\ (\ref{keplerian_rl}) for spinless black holes, i.e., for $a = 0$ was derived already in \cite{mm2015}, basing on the post-Newtonian expansion. It reads
\begin{equation}
 j(\Omega) = \left(- 3 \Omega + w^{-4/3} \Omega^{1/3} \right)^{-1}.
\end{equation}
Trying to apply this rotation law to the case with $a \neq 0$, one finds a mass gap for $a < 0$ and $\Omega > 0$, i.e., for counter-rotating disks---numerical solutions can be found only if the mass of the disk is sufficiently large \cite{kkmmop2}. A similar effect has been observed for co-rotating disks in \cite{kmm2021}, for yet another rotation law. In summary, putting a strong, arbitrarily distributed magnetic field is also sufficient to remove the exceptional property of the Keplerian rotation law (\ref{keplerian_rl})---allowing for compact and light disks.

The effect of the self-gravity of the disk on the location of the ISCO is small for low-mass disks. It depends both on the mass of the disk and on the location of its inner edge $R_1$. In general, the ISCO radius $R_\mathrm{c,ISCO}$ grows with the increasing mass of the disk. To make this effect more visible, we provide additional data in Table \ref{tab4}. Solutions collected in Table \ref{tab4} were computed assuming smaller inner disk radii (comparable to ISCO radii) and an increased spatial resolution of the numerical grid.

\begin{table*}
\caption{\label{tab1} Parameters of numerical solutions corresponding to magnetized and non-magnetized disks. Here the coordinate inner and outer radii of the disk are fixed: $R_{1}=5$, $R_{2}=20$. The polytropic exponent is $\Gamma=4/3$. The black hole mass parameter $m = 1$.}
\begin{ruledtabular}
\begin{tabular}{ccccccccccc}
$n$	&	$a$	&	$10^4\rho_\mathrm{max}$	&	$C_1$	&	$m_\text{ADM}$ &$M_{\mathrm{BH}}$ & $\beta_\text{mag}$ & $R_{\mathrm{c}1}$	& $R_{\mathrm{c}2}$	 & $r_\text{ISCO}$ & $r_{\mathrm{c,ISCO}}$	 \\\hline
2	&	0.9	&	3.5	&	0	&	1.58    & 1.01  	&	$\infty $ & 6.45  & 21.78 &	1.29 & 2.70	\\
	&		&		&	0.1	&	1.82	& 1.01      &	2.20   	  & 6.60  & 22.06 &	1.29 & 2.73	\\\hline													
2	&	0.9	&	2.5	&	0	&	1.34	& 1.00      &  $\infty $  &	6.31  & 21.50 & 1.29 & 2.67	\\
	&		&		&	0.1	&	1.49	& 1.00      &	1.84	  & 6.40  & 21.67 & 1.29 & 2.69	\\
	&		&		&	0.2	&	1.56	& 1.01      &	0.95	  &	6.45  & 21.75 & 1.29 & 2.70 \\
	&		&		&	0.3	&	1.60	& 1.01      &	0.59	  &	6.48  & 21.80 & 1.29 & 2.70	\\
	&		&		&	0.4	&	1.63	& 1.01      &	0.39	  &	6.51  & 21.83 & 1.29 & 2.71	\\
	&		&		&	0.5	&	1.64	& 1.01      & 	0.24	  & 6.52  &	21.84 & 1.29 & 2.71	\\
	&		&		&	0.6	&	1.65	& 1.01      &   0.14	  &	6.53  & 21.85 & 1.29 & 2.71	\\
	&		&		&	0.7	&	1.65	& 1.01  	&   0.05	  &	6.54  & 21.85 & 1.29 & 2.71	\\\hline													
2	&	0.9	&	1	&	0	&	1.08	& 1.00      & $\infty $	  & 6.17  & 21.20 &	1.29 & 2.64	\\
	&		&		&	0.1	&	1.12	& 1.00      &	1.15	  & 6.19  & 21.25 &	1.29 & 2.65	\\
	&		&		&	0.2	&	1.14	& 1.00      &	0.50	  & 6.21  & 21.26 & 1.29 & 2.65	\\
	&		&		&	0.3	&	1.14	& 1.00      &	0.23	  & 6.21  & 21.26 &	1.29 & 2.65	\\
	&		&		&	0.4	&	1.14	& 1.00      &	0.06	  & 6.21  & 21.26 &	1.29 & 2.65	\\\hline													
2	&	0.5	&	2.5	&	0	&	1.36	& 1.02      & $\infty $	& 6.30 & 21.50  &	3.31 & 4.54	\\
	&		&		&	0.1	&	1.51	& 1.03      &	1.90	& 6.40 & 21.67  &	3.34 & 4.64	\\
	&		&		&	0.2	&	1.58	& 1.03      &	0.97	& 6.45 & 21.75  &	3.42 & 4.76	\\
	&		&		&	0.3	&	1.62	& 1.04      &	0.61	& 6.48 & 21.80  &	3.42 & 4.78	\\
	&		&		&	0.4	&	1.65	& 1.04      &	0.40	& 6.51 & 21.83  &	3.46 & 4.83	\\\hline													
2	&	0.5	&	1	&	0	&	1.08	& 1.00      &  $\infty$ & 6.14 & 21.18  &	3.2	& 4.33 \\
	&		&		&	0.1	&	1.13	& 1.01      &	1.15	& 6.16 & 21.23  &	3.23 & 4.38	\\
	&		&		&	0.2	&	1.14	& 1.01      &	0.51	& 6.18 & 21.24  &	3.23 & 4.39	\\
	&		&		&	0.3	&	1.14	& 1.01      &	0.23	& 6.18 & 21.25  &	3.23 & 4.39	\\
	&		&		&	0.4	&	1.14	& 1.01      &	0.06	& 6.18 & 21.25  &	3.23 & 4.39	\\\hline
1.5	&	0.9	&	3.5	&	0.0	&	1.58	& 1.00 & $\infty $	&	6.45	&	21.78	&	1.29	&	2.70	\\
	&		&		&	0.1	&	1.78	&	1.01	&	2.82	&	6.57	&	22.01	&	1.29	&	2.72	\\\hline
																					
1.5	&	0.9	&	2.5	&	0.0	&	1.34	& 1.00	&	$\infty $	& 6.31	&	21.50	&	1.29	&	2.67	\\
	&		&		&	0.1	&	1.46	&	1.00	&	2.38	&	6.39	&	21.64	&	1.29	&	2.68	\\
	&		&		&	0.2	&	1.53	&	1.00	&	1.27	&	6.43	&	21.71	&	1.29	&	2.69	\\
	&		&		&	0.3	&	1.57	&	1.00	&	0.83	&	6.46	&	21.76	&	1.29	&	2.70	\\
	&		&		&	0.4	&	1.56	&	1.01	&	0.59	&	6.48	&	21.80	&	1.29	&	2.70	\\
	&		&		&	0.5	&	1.62	&	1.01	&	0.43	&	6.50	&	21.82	&	1.29	&	2.71	\\
	&		&		&	0.6	&	1.63	&	1.01	&	0.31	&	6.51	&	21.83	&	1.29	&	2.71	\\
	&		&		&	0.7	&	1.64	&	1.01	&	0.22	&	6.52	&	21.84	&	1.29	&	2.71	\\
	&		&		&	0.8	&	1.65	&	1.01	&	0.14	&	6.53	&	21.85	&	1.29	&	2.71	\\
	&		&		&	0.9	&	1.65	&	1.01	&	0.07	&	6.53	&	21.85	&	1.29	&	2.71	\\
	&		&		&	1.0	&	1.65	&	1.01	&	0.02	&	6.54	&	21.85	&	1.29	&	2.71	\\			
\end{tabular}
\end{ruledtabular}
\end{table*}

\begin{table*}
\caption{\label{tab2} Same as in Table \ref{tab1}. Here $R_1=3$, $R_2=20$, $a=0.9$, $\Gamma=4/3$ $C_1=1.0$, and $m = 1$.}
\begin{ruledtabular}
\begin{tabular}{ccccccccc}
        $n$ & $10^4 \rho_\mathrm{max}$ & $M_{\mathrm{BH}}$ & $m_{\mathrm{ADM}}$ & $\beta_{\mathrm{mag}}$ &$R_{\mathrm{c}1}$ & $R_{\mathrm{c}2}$& $r_{\mathrm{ISCO}}$ & $r_{\mathrm{c,ISCO}}$ \\
        \hline
         2 & 3.0 & 1.003 & 1.333 & 0.158 & 4.328 & 21.474 & 1.294 & 2.686 \\
         2 & 2.75 & 1.003 & 1.273 & 0.102 & 4.300 & 21.407 & 1.294 & 2.676 \\
         2 & 2.5 & 1.002 & 1.219 & 0.0422 & 4.275 & 21.347 & 1.294 & 2.668 \\
         2 & 2.4 & 1.002 & 1.199 & 0.0177 & 4.266 & 21.325 & 1.294 & 2.665 \\\hline
         1.5 & 2.4 & 1.002 & 1.240 & 0.265 & 4.280 & 21.371 & 1.294 & 2.670 \\
         1.5 & 2.2 & 1.002 & 1.199 & 0.210 & 4.261 & 21.325 & 1.294 & 2.664 \\
         1.5 & 2.0 & 1.002 & 1.162 & 0.151 & 4.244 & 21.283 & 1.294 & 2.659 \\
         1.5 & 1.8 & 1.001 & 1.128 & 0.0868 & 4.229 & 21.245 & 1.294 & 2.653 \\
         1.5 & 1.6 & 1.001 & 1.097 & 0.0133 & 4.215 & 21.211 & 1.294 & 2.648 \\\hline
         1 & 2.5 & 1.003 & 1.303 & 0.671 & 4.302 & 21.441 & 1.294 & 2.678  \\
         1 & 1.5 & 1.001 & 1.111 & 0.335 & 4.218 & 21.227 & 1.294 & 2.650 \\
         1 & 1.2 & 1.001 & 1.069 & 0.190 & 4.199 & 21.179 & 1.294 & 2.644 \\
         1 & 0.95 & 1.000 & 1.039 & 0.0328 & 4.186 & 21.146 & 1.294 & 2.640 \\
         1 & 0.92 & 1.000 & 1.036 & 0.0104 & 4.185 & 21.142 & 1.294 & 2.639 \\\hline
         0.5 & 1.5 & 1.001 & 1.138 & 1.07 & 4.225 & 21.257 & 1.294 & 2.653 \\ 
         0.5 & 0.85 & 1.000 & 1.048 & 0.616 & 4.188 & 21.156 & 1.294 & 2.640 \\
         0.5 & 0.55 & 1.000 & 1.019 & 0.294 & 4.175 & 21.123 & 1.294 & 2.637
\end{tabular}
\end{ruledtabular}
\end{table*}

\begin{table*}
\caption{\label{tab3} Same as in Table \ref{tab1}. Here $R_{1}=3$, $R_{2}=20$, $a=0.9$, $\Gamma=4/3$ $C_1=0.5$, and $m = 1$.}
\begin{ruledtabular}
\begin{tabular}{ccccccccc}
       $n$ & $10^4 \rho_\mathrm{max}$ & $M_{\mathrm{BH}}$ & $m_{\mathrm{ADM}}$ & $\beta_{\mathrm{mag}}$ &$R_{\mathrm{c}1}$ & $R_{\mathrm{c}2}$& $r_{\mathrm{ISCO}}$ & $r_{\mathrm{c,ISCO}}$ \\
       \hline
 2 & 1.03 & 1.000 & 1.050 & 0.096 & 4.213 & 21.219 & 1.285 & 2.634 \\
 1 & 0.85 & 1.000 & 1.050 & 0.629 & 4.211 & 21.219 & 1.285 & 2.633 \\
 0.5 & 0.79 & 1.000 & 1.050 & 1.373 & 4.210 & 21.219 & 1.285 & 2.633 \\
  - & 0.84 & 1.000 & 1.050 & $\infty$ & 4.208 & 21.219 & 1.285 & 2.633 \\
  \hline
 2   & 1.76 & 1.001 & 1.158 & 0.441 & 4.261 & 21.339 & 1.298 & 2.660 \\
 1   & 1.60 & 1.001 & 1.158 & 1.123 & 4.256 & 21.340 & 1.298 & 2.659 \\
 0.5 & 1.56 & 1.001 & 1.158 & 2.213 & 4.253 & 21.340 & 1.298 & 2.658 \\
  - & 1.67 & 1.001 & 1.158 & $\infty$ & 4.249 & 21.341 & 1.285 & 2.647 \\
  \hline
 2 & 2.35 & 1.003 & 1.276 & 0.631 & 4.313 & 21.472 & 1.298 & 2.678 \\
 1 & 2.21 & 1.003 & 1.276 & 1.408 & 4.306 & 21.473 & 1.298 & 2.676 \\
 0.5 & 2.21 & 1.003 & 1.276 & 2.722 & 4.302 & 21.473 & 1.298 & 2.675 \\
  - & 2.36 & 1.003 & 1.276 & $\infty$ & 4.296 & 21.474 & 1.298 & 2.673 \\
  \hline
 2 & 2.89 & 1.004 & 1.407 & 0.780 & 4.371 & 21.619 & 1.298 & 2.697 \\
 1 & 2.78 & 1.004 & 1.407 & 1.659 & 4.362 & 21.620 & 1.298 & 2.695 \\
 0.5 & 2.80 & 1.004 & 1.407 & 3.163 & 4.357 & 21.620 & 1.298 & 2.693 \\
  - & 2.99 & 1.004 & 1.407 & $\infty$ & 4.349 & 21.621 & 1.298 & 2.692 \\
  \hline
 2 & 3.39 & 1.005 & 1.551 & 0.931 & 4.436 & 21.781 & 1.298 & 2.719 \\
 1 & 3.32 & 1.005 & 1.551 & 1.870 & 4.425 & 21.782 & 1.298 & 2.716 \\
 0.5 & 3.36 & 1.005 & 1.551 & 3.555 & 4.418 & 21.783 & 1.298 & 2.715 \\
  - & 3.59 & 1.005 & 1.551 & $\infty$ & 4.410 & 21.784 & 1.298 & 2.712 \\
  \hline
2 & 4.11 & 1.008 & 1.796 & 1.127 & 4.549 & 22.056 & 1.298 & 2.758 \\
1 & 4.08 & 1.008 & 1.796 & 2.217 & 4.535 & 22.058 & 1.298 & 2.754 \\
0.5 & 4.17 & 1.008 & 1.796 & 4.190 & 4.527 & 22.059 & 1.298 & 2.752 \\
 - & 4.45 & 1.008 & 1.796 & $\infty$ & 4.516 & 22.061 & 1.298 & 2.749 \\
 \hline
 2 & 4.56 & 1.010 & 1.980 & 1.258 & 4.636 & 22.264 & 1.298 & 2.788 \\
 1 & 4.57 & 1.010 & 1.980 & 2.447 & 4.620 & 22.266 & 1.298 & 2.784 \\
 0.5 & 4.69 & 1.010 & 1.980 & 4.562 & 4.611 & 22.268 & 1.298 & 2.781 \\
  - & 5.00 & 1.010 & 1.980 & $\infty$ & 4.599 & 22.270 & 1.298 & 2.778 \\
\end{tabular}
\end{ruledtabular}
\end{table*}

\begin{table*}
\caption{\label{tab4} Same as in Table \ref{tab1}. Solutions collected in this table were obtained with a larger numerical grid. The spatial grid
resolution around the ISCO is $\Delta r \approx 0.01$. Here: $R_{1}=1.5$, $R_{2}=20$, $a=0.9$, $\Gamma=4/3$, $C_1=1.0$, and $m = 1$.}
\begin{ruledtabular}
\begin{tabular}{ccccccccc}
        $n$ & $10^4 \rho_\mathrm{max}$ & $M_{\mathrm{BH}}$ & $m_{\mathrm{ADM}}$ & $\beta_{\mathrm{mag}}$ &$R_{\mathrm{c}1}$ & $R_{\mathrm{c}2}$& $r_{\mathrm{ISCO}}$ & $r_{\mathrm{c,ISCO}}$ \\
        \hline
         2 & 6.0 & 1.002 & 1.272 & 0.988 & 2.891 & 21.361 & 1.312 & 2.713\\
         2 & 5.0 & 1.002 & 1.159 & 0.688 & 2.858 & 21.238 & 1.302 & 2.678\\
         2 & 4.0 & 1.001 & 1.076 & 0.333 & 2.834 & 21.147 & 1.292 & 2.651\\\hline
         1 & 6.0 & 1.005 & 1.443 & 2.674 & 2.932 & 21.548 & 1.312 & 2.747\\
         1 & 5.0 & 1.003 & 1.299 & 2.216 & 2.891 & 21.391 & 1.302 & 2.705\\
         1 & 4.0 & 1.002 & 1.182 & 1.779 & 2.859 & 21.263 & 1.302 & 2.680\\
         1 & 3.0 & 1.001 & 1.090 & 1.249 & 2.834 & 21.163 & 1.292 & 2.652
 \end{tabular}
\end{ruledtabular}
\end{table*}

\section{Massive disks: Bifurcations}

\begin{figure}
    \includegraphics[width=\columnwidth]{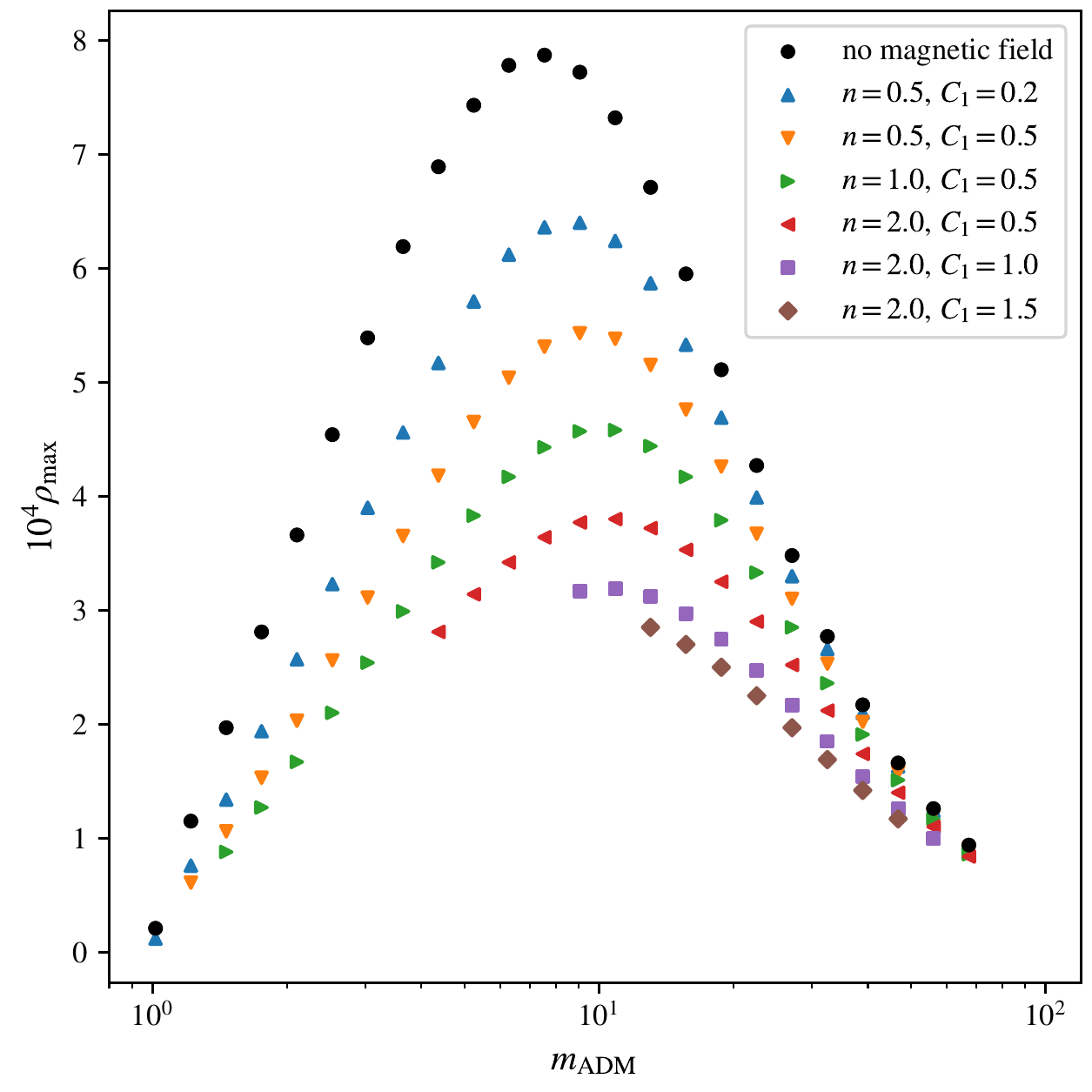}
    \caption{Maximum rest-mass density within the disk $10^4 \rho_\mathrm{max}$ vs.\ the total asymptotic mass $m_\mathrm{ADM}$. Here $m = 1$, $a = 0.9$, $\Gamma = 4/3$, $R_1 = 10$, $R_2 = 25$. Different colors correspond to configurations with different values of $n$ and $C_1$. Black dots depict solutions with no magnetic field.}
    \label{bif1}
\end{figure}

\begin{figure}
    \includegraphics[width=\columnwidth]{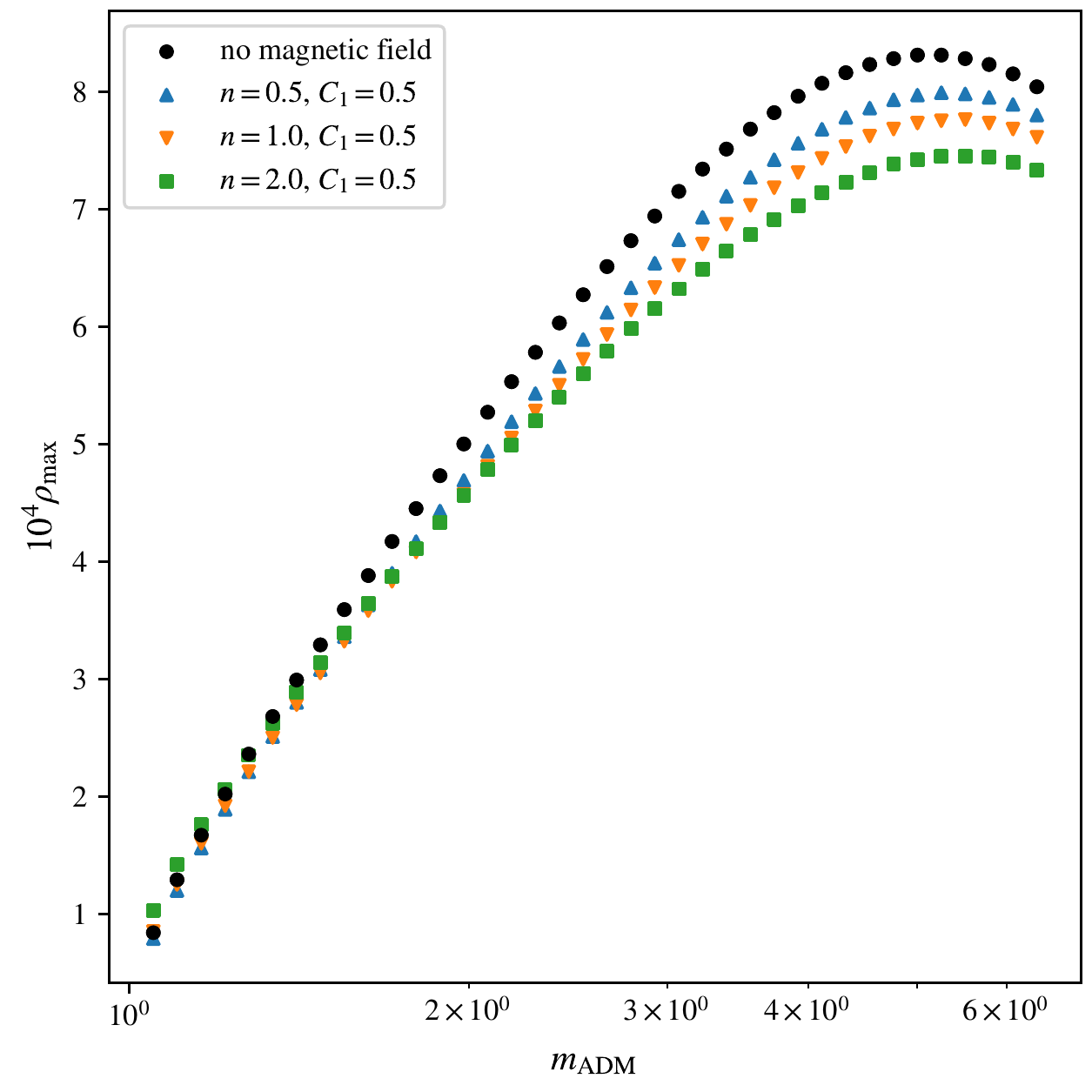}
    \caption{Maximum rest-mass density within the disk $10^4 \rho_\mathrm{max}$ vs.\ the total asymptotic mass $m_\mathrm{ADM}$. Here $m = 1$, $a = 0.9$, $\Gamma = 4/3$, $R_1 = 3$, $R_2 = 20$. Different colors correspond to configurations with different values of $n$ and $C_1$. Black dots depict solutions with no magnetic field.}
    \label{bif2}
\end{figure}

\begin{figure}
    \includegraphics[width=\columnwidth]{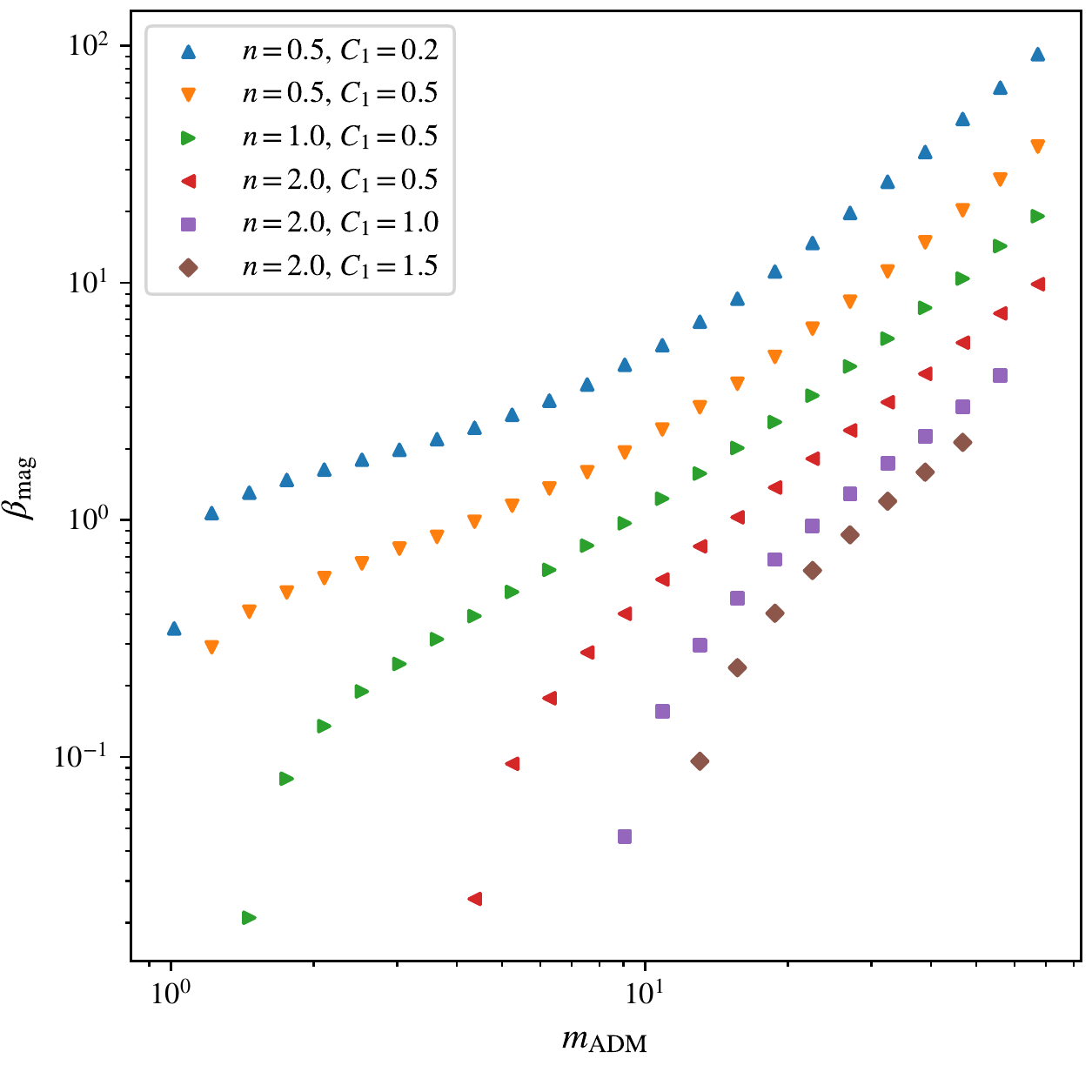}
    \caption{Magnetization parameter $\beta_\mathrm{mag}$ vs.\ the total asymptotic mass $m_\mathrm{ADM}$. The data are the same as in Fig.\ \ref{bif1}, except for solutions with no magnetic field, which have not been plotted.}
    \label{bif1bmag}
\end{figure}

\begin{figure}
    \includegraphics[width=\columnwidth]{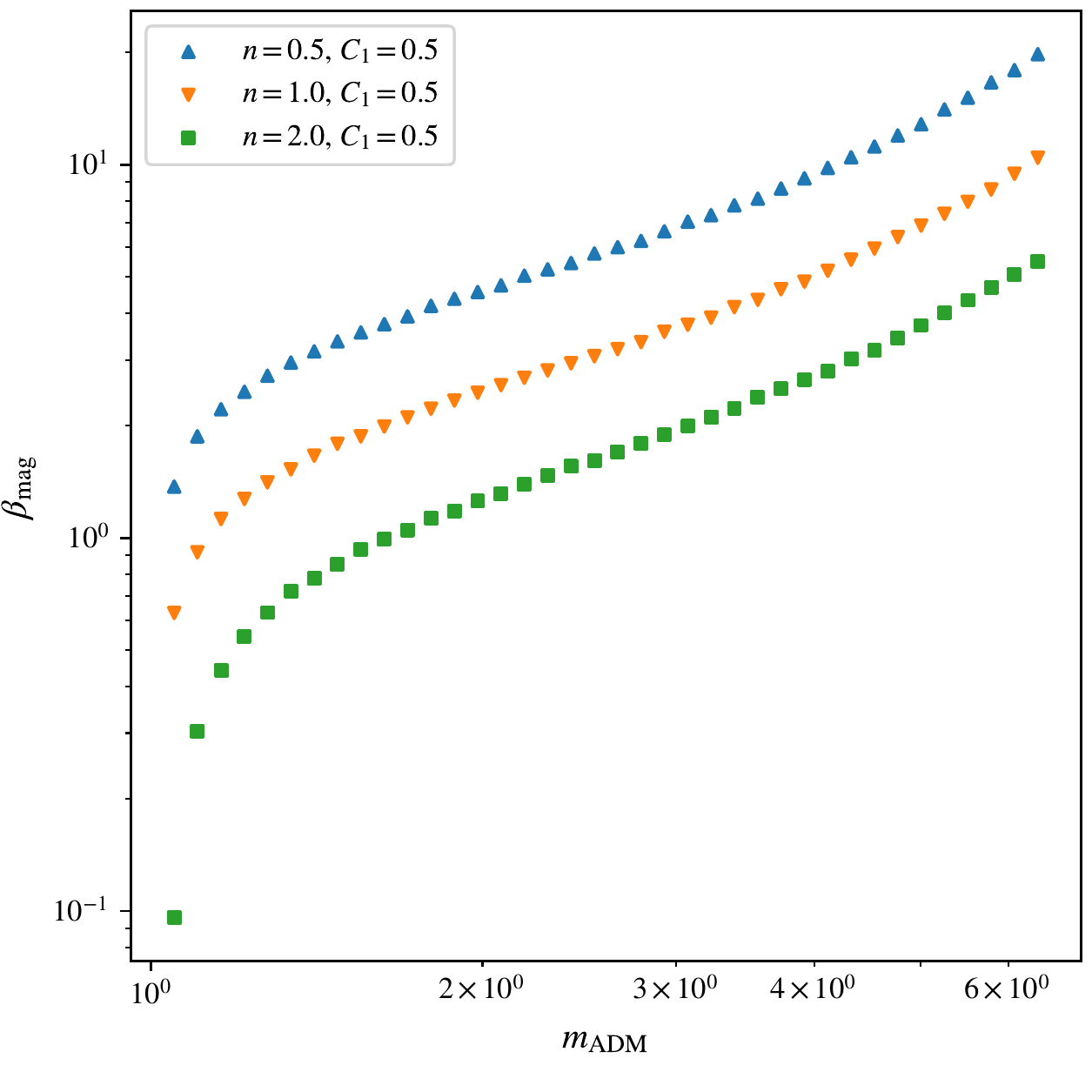}
    \caption{Magnetization parameter $\beta_\mathrm{mag}$ vs.\ the total asymptotic mass $m_\mathrm{ADM}$. The data are the same as in Fig.\ \ref{bif2}, except for solutions with no magnetic field, which have not been plotted.}
    \label{bif2bmag}
\end{figure}

\begin{figure}
    \includegraphics[width=\columnwidth]{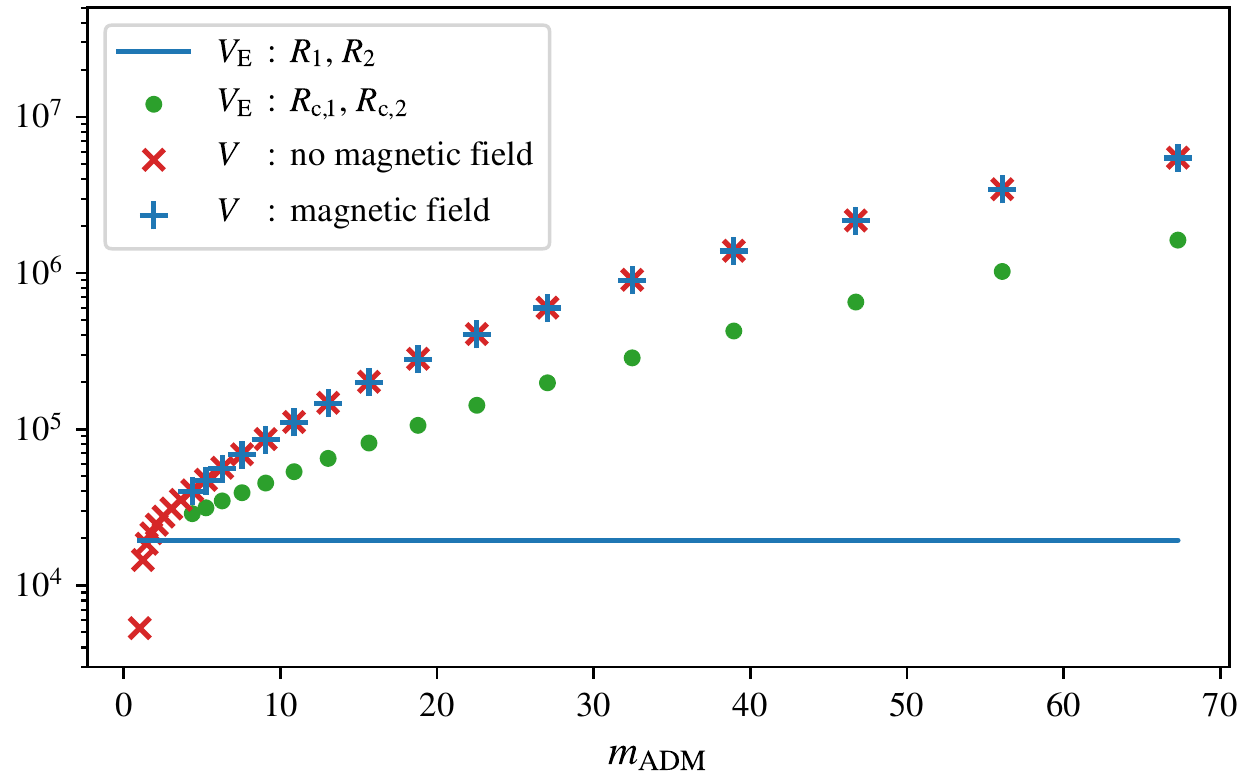}
    \caption{Disk volumes vs.\ the asymptotic mass $m_\mathrm{ADM}$. Blue  crosses and green dots correspond to magnetized disks with $n = 2$ and $C_1 = 0.5$. Blue crosses depict the proper volume $V$ computed according to Eq.\ (\ref{propervolume}). Green dots correspond to Euclidean volumes $V_\mathrm{E}$ computed according to Eq.\ (\ref{euclideanvolume}) using circumferential inner and outer radii of the disk $R_\mathrm{c,1}$ and $R_\mathrm{c,2}$ in place of $R_1$ and $R_2$. Red crosses (overlapping with blue ones) depict proper volumes $V$ of disks with no magnetic fields. In all models $m = 1$, $a = 0.9$, $R_1 = 10$, $R_2 = 25$, $\Gamma = 4/3$. The Euclidean volume $V_\mathrm{E}$ computed according to Eq.\ (\ref{euclideanvolume}) using coordinate radii $R_1$ and $R_2$ is shown with a blue line.}
    \label{bif1dvol}
\end{figure}

\begin{figure}
    \includegraphics[width=\columnwidth]{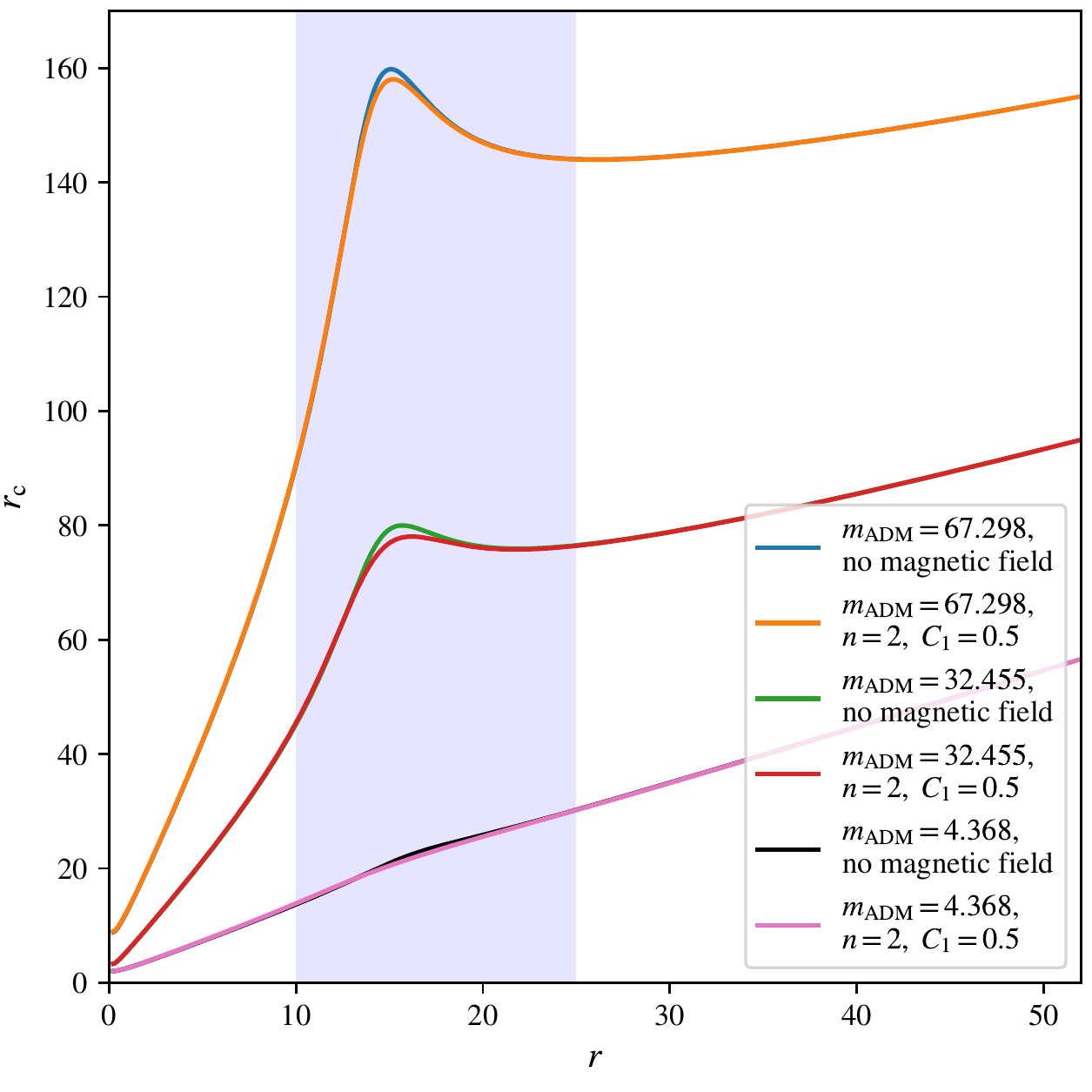}
    \caption{Circumferential radius $r_\mathrm{c}$ vs.\ the coordinate radius $r$ at the equatorial plane. We plot pairs of solutions characterised by asymptotic masses (from top to bottom) $m_\mathrm{ADM} = 67.298$, 32.455, and 4.368. Each pair consists of a solution corresponding to a magnetized disk and a model with no magnetic field. The parameters common to all six solutions are $m = 1$, $a = 0.9$, $\Gamma = 4/3$, $R_1 = 10$, $R_2 = 25$. Magnetized disks are obtained assuming $n = 2$ and $C_1 = 0.5$. They are characterized by magnetization parameters (from top to bottom) $\beta_\mathrm{mag} = 9.876$, 3.137, and 0.0252, respectively. The region corresponding to the disk---between $R_1$ and $R_2$---is marked in pale blue color.}
    \label{rc1}
\end{figure}

\begin{figure}
    \includegraphics[width=\columnwidth]{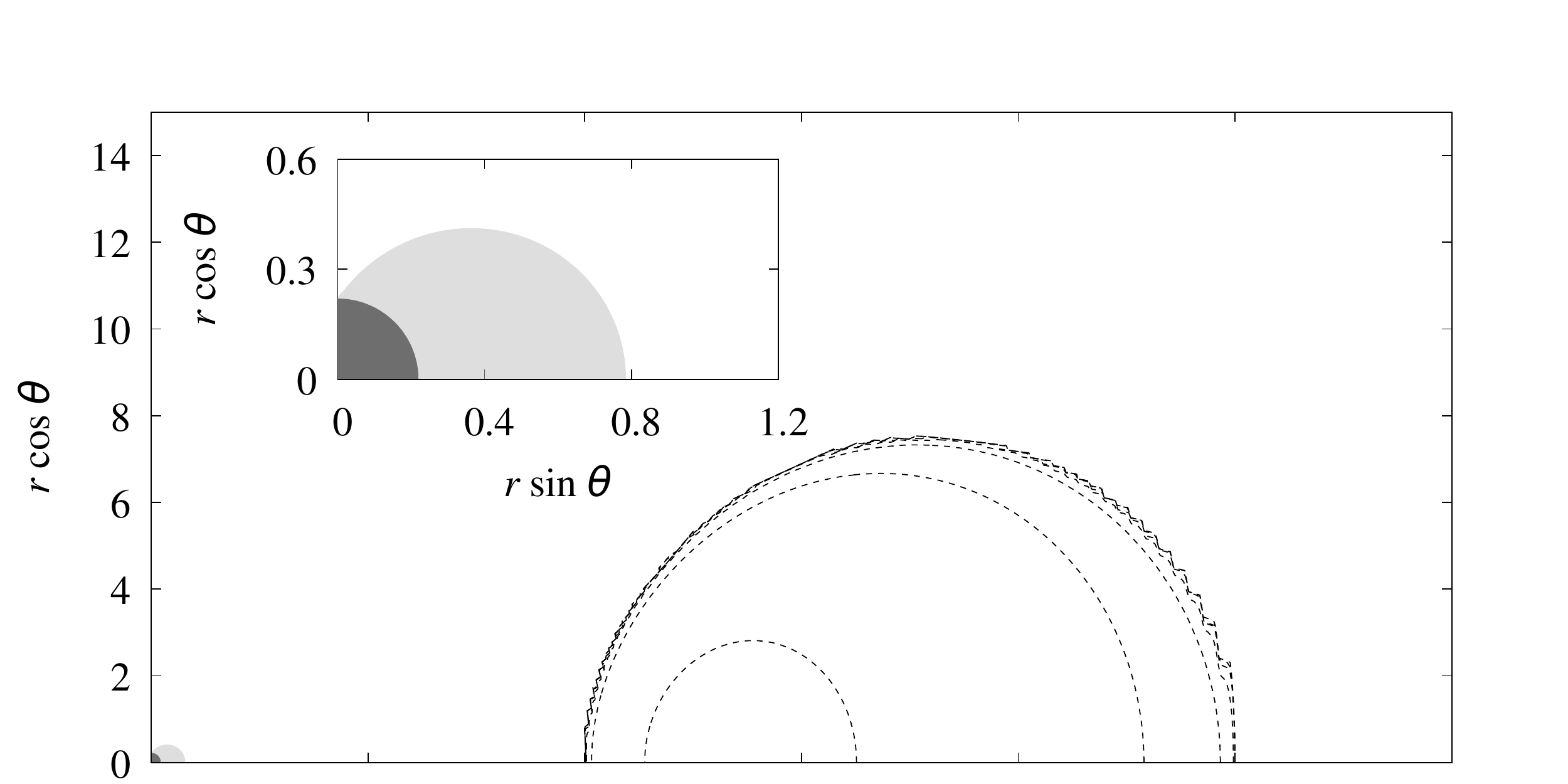}\\
    \includegraphics[width=\columnwidth]{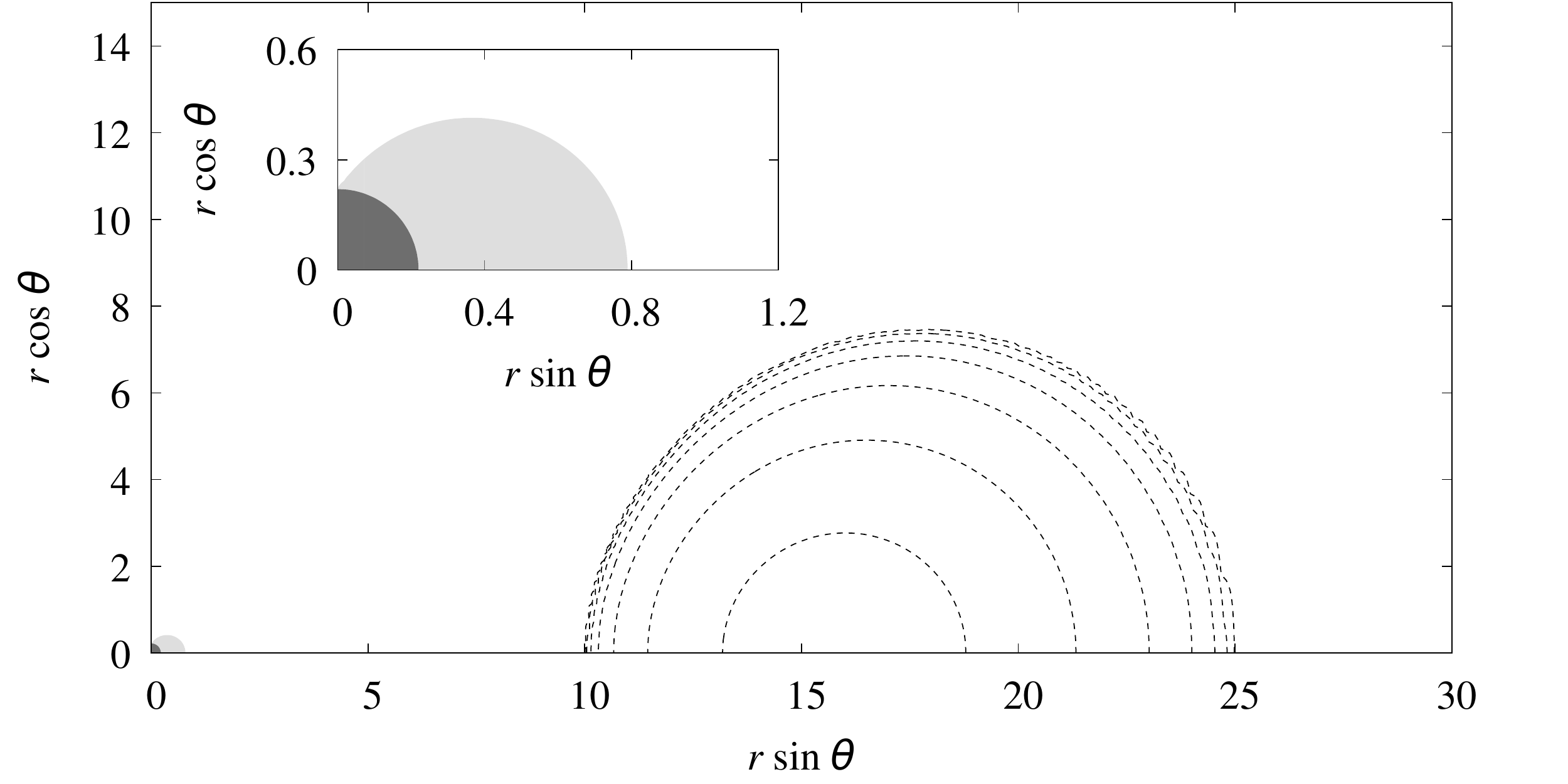}
    \caption{Black hole ergoregions for two configurations with $m = 1$, $a = 0.9$, $\Gamma = 4/3$, $R_1 = 10$, $R_2 = 25$, $m_\mathrm{ADM} = 4.368$. Ergoregions are marked in gray. Contours of constant rest-mass density correspond to $\rho = 2 \times 10^i$, where $i = -10, \dots, -4$. Upper panel: $n = 2$, $C_1 = 0.5$, $\beta_\mathrm{mag} = 0.0252$. Lower panel: no magnetic field.}
    \label{profrhoErgbp}
\end{figure}

\begin{figure}
    \includegraphics[width=\columnwidth]{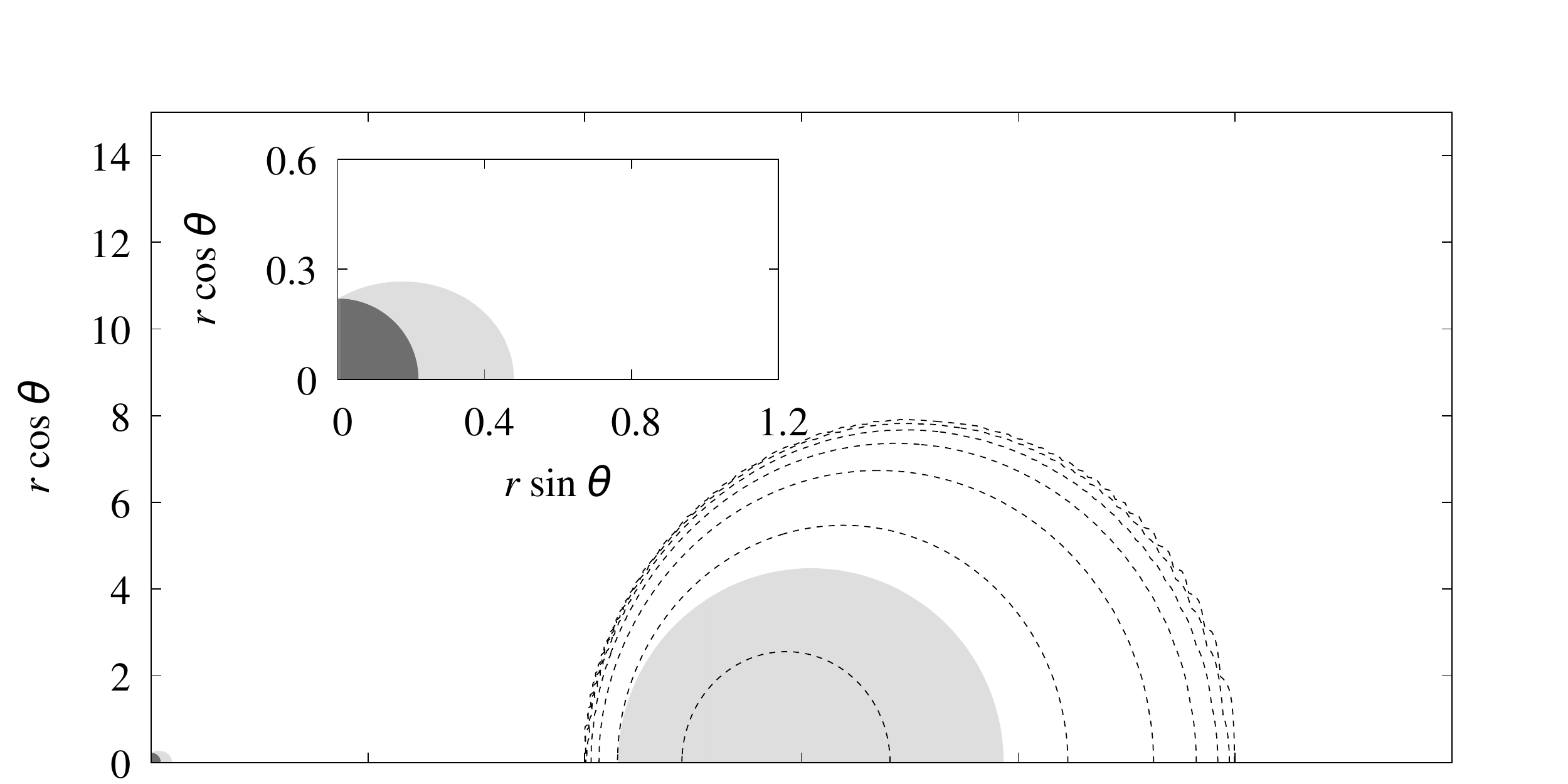}\\
    \includegraphics[width=\columnwidth]{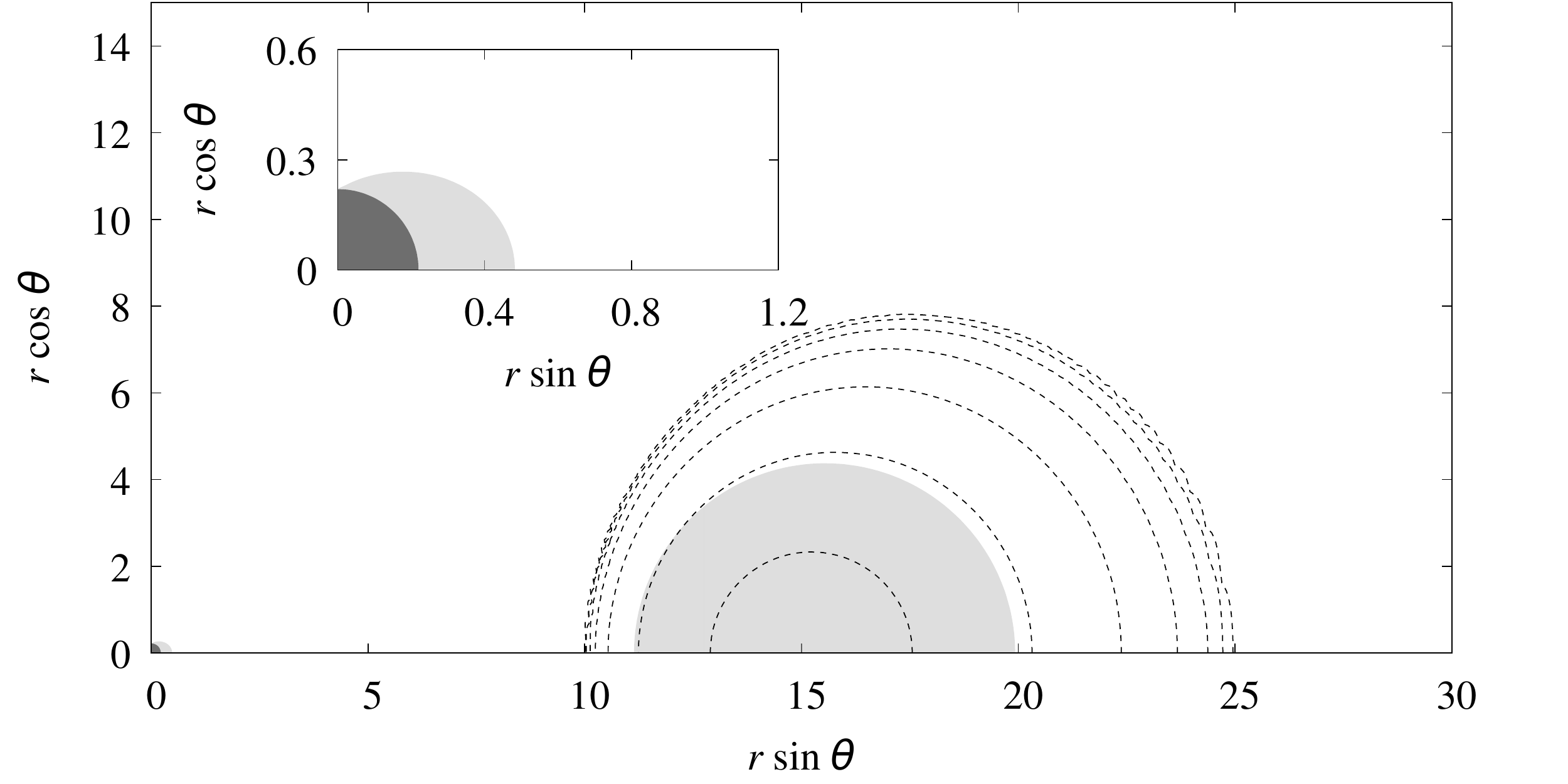}
    \caption{Ergoregions in two configurations with $m = 1$, $a = 0.9$, $\Gamma = 4/3$, $R_1 = 10$, $R_2 = 25$, $m_\mathrm{ADM} = 15.651$. Ergoregions are marked in gray. Contours of constant rest-mass density correspond to $\rho = 2 \times 10^i$, where $i = -10, \dots, -4$. Upper panel: $n=2$, $C_1=0.5$, $\beta_\mathrm{mag}=1.026$. Lower panel: no magnetic field.}
    \label{profrhoErgbp2}
\end{figure}

\begin{figure}
    \includegraphics[width=\columnwidth]{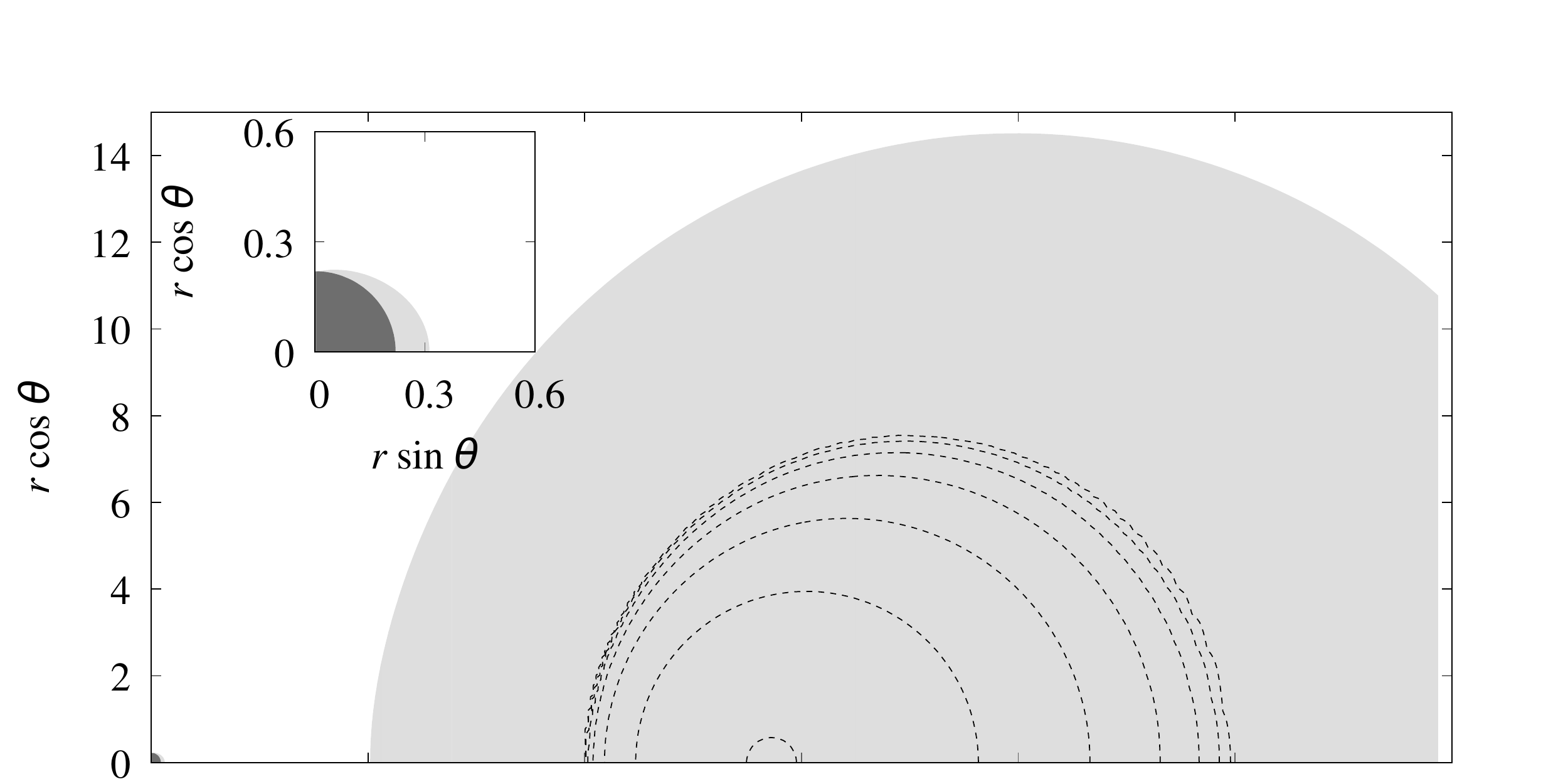}\\
    \includegraphics[width=\columnwidth]{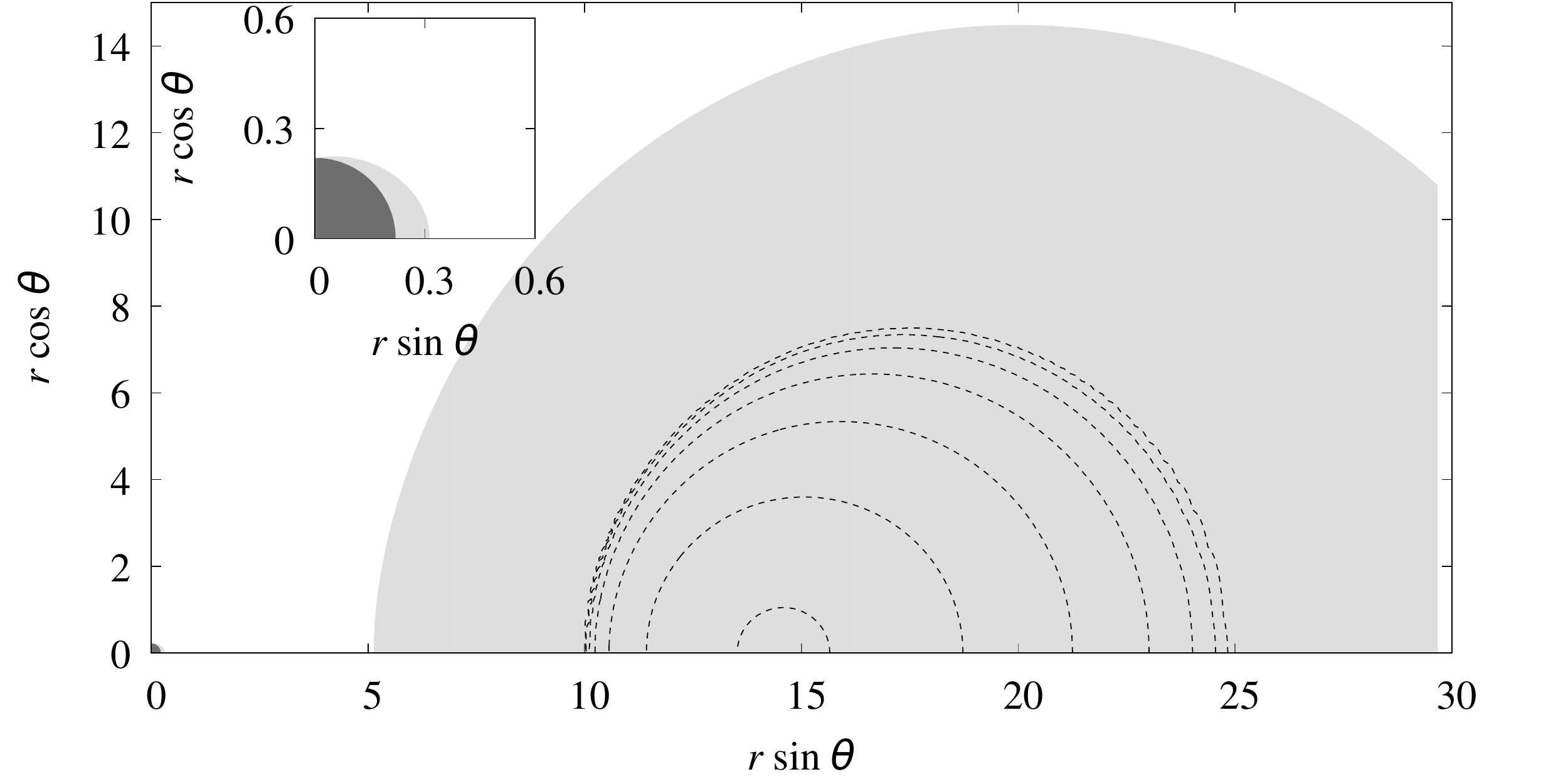}
    \caption{Ergoregions in two configurations with $m = 1$, $a = 0.9$, $\Gamma = 4/3$, $R_1 = 10$, $R_2 = 25$, $m_\mathrm{ADM} = 32.455$. Ergoregions are marked in gray. Contours of constant rest-mass density correspond to $\rho = 2 \times 10^i$, where $i = -10, \dots, -4$. Upper panel: $n=2$, $C_1=0.5$, $\beta_\mathrm{mag} = 3.137$. Lower panel: no magnetic field.}
    \label{profrhoErgbp3}
\end{figure}

Much more interesting features can be observed for massive disks. Figures \ref{bif1} and \ref{bif2} show bifurcation diagrams obtained for magnetized configurations with $m = 1$, $\Gamma = 4/3$, $a = 0.9$, and $R_1 = 10$, $R_2 = 25$ (Fig.\ \ref{bif1}), or $R_1 = 3$, $R_2 = 20$ (Fig.\ \ref{bif2}). They are analogous to the ones shown in \cite{ergosfery} for non magnetized disks. Each point on these diagrams corresponds to a different stationary solution. We plot the value of the maximum of the rest-mass density within the disk $\rho_\mathrm{max}$ vs.\ the total asymptotic mass of the system $m_\mathrm{ADM}$. For a given set of parameters $m$, $\Gamma$, $a$, $R_1$, $R_2$, and $\rho_\mathrm{max}$, there usually exist two solutions, differing in the asymptotic mass of the system. Moreover, there is a limit on the allowed value of the maximum rest-mass density within the disk, above which no solutions are found. The fact that two solutions with different masses can be characterized by similar geometric parameters describing the shape of the torus and the same maximal rest-mass density can be explained by a difference in the proper volume of the torus. This is illustrated in Fig.\ \ref{bif1dvol}, in which we plot the volume of the tori corresponding to a selection of solutions shown in Fig.\ \ref{bif1}. A behavior of this kind is known in general-relativistic systems, mostly in spherical symmetry \cite{kmmmx,bizon}. We were quite surprised to observe it for non-magnetized disk--black hole systems in \cite{ergosfery}. Due to axial symmetry, this effect was referred to as the breaking of the Pappus-Guldinus rule---the proper volume of the torus turns out to be significantly different than its estimate based on external characteristics of the torus (inner and outer circumferences) and Euclidean formulas. The Euclidean (Pappus-Guldinus) formula for the volume of the torus with the inner and outer radii equal to $R_1$ and $R_2$, respectively, reads
\begin{equation}
\label{euclideanvolume}
 V_\mathrm{E} = \frac{\pi^2}{4} (R_1 + R_2)^2(R_2 - R_1)^2.
\end{equation}
In the general-relativistic context, formula (\ref{euclideanvolume}) can be used either with coordinate radii $R_1$ and $R_2$ or with geometric circumferential radii $R_\mathrm{c,1}$, $R_\mathrm{c,2}$. Both versions of this volume estimate are plotted in Fig.\ \ref{bif1dvol}, together with the true geometric volume given by
\begin{equation}
\label{propervolume}
V = 2 \pi \int dr \int d \theta r^2 \sin \theta \psi^6  e^{2 q},
\end{equation}
where the integral is performed over the disk region. Of course, the surfaces of our disks are not perfect geometric tori. To some extent, the comparison between $V$ and $V_\mathrm{E}$ is justified by the shape of massive disks, whose meridional cross sections appear to be roughly circular. This shape is illustrated in Figs.\ \ref{profrhoErgbp}--\ref{profrhoErgbp3}. By inspecting Fig.\ \ref{bif1dvol} we see that the true geometrical volume of massive disks shown in Fig.\ \ref{bif1} is much larger than its Euclidean estimates $V_\mathrm{E}$. 

Figure \ref{bif1bmag} shows the magnetization parameter $\beta_\mathrm{mag}$ for the solutions depicted in Fig.\ \ref{bif1} (except for non-magnetized configurations, for which $\beta_\mathrm{mag}= \infty$). Clearly, $\beta_\mathrm{mag}$ grows with increasing $m_\mathrm{ADM}$. Thus, in practice, our most massive disks are also relatively weakly magnetized. A comparison of Figs.\ \ref{bif1} and \ref{bif1bmag} shows that the values of the maximum density in the disk $\rho_\mathrm{max}$ essentially grows with the magnetization parameter $\beta_\mathrm{mag}$ for both bifurcation branches, i.e., both for massive and light tori. The upper limit on $\rho_\mathrm{max}$ corresponds to the non-magnetized case. Also, the asymptotic mass $m_\mathrm{ADM}$ corresponding to the critical solution with the maximal rest-mass density in the torus $\rho_\mathrm{max}$ grows with the magnetization parameter $\beta_\mathrm{mag}$. The former can change for a different choice of inner and outer disk radii $R_1$ and $R_2$, as illustrated in Fig.\ \ref{bif2}. In this case the disks originate much closer to the central black hole. For light disks the dependence of the maximum rest-mass density $\rho_\mathrm{max}$ on the magnetization parameter can be reversed---less magnetized disks can be characterized by smaller values of $\rho_\mathrm{max}$ for the same value of $m_\mathrm{ADM}$.

Note that the volume of the torus $V$ depends weekly on the magnetization parameter, and it is essentially a function of the total mass $m_\mathrm{ADM}$ (Fig.\ \ref{bif1dvol}). On the other hand, the maximal density within the disk corresponding to a fixed $m_\mathrm{ADM}$ depends quite strongly on magnetization, as shown in Fig.\ \ref{bif1}. This is consistent with the observation that the the magnetic field mainly affects the distribution of matter, and its effect on the spacetime geometry is indirect, through the distribution of mass.

Another effect illustrated in Fig.\ \ref{bif1} is the mass gap occurring for highly magnetized disks. As remarked in Sec.\ \ref{lowmass}, numerical solutions corresponding to highly magnetized disks can only be found if the mass of the disks (or the total asymptotic mass) is sufficiently high. As a result, lower (more magnetized) branches in Fig.\ \ref{bif1} originate at quite high ADM masses.

Another way to visualise the space time geometry within the disk is to look at the relation between the geometric circumferential radius $r_\mathrm{c}$ and the coordinate radius $r$, say at the equatorial plane. In general this relation does not have to be monotonic, and consequently $r_\mathrm{c}$ would be a bad candidate for a coordinate. The fact that the circumferential radius $r_\mathrm{c}$ can have a local maximum within the disk was observed for the non-magnetized case in \cite{labranche, meinel} and \cite{ergosfery}. Putting this in more picturesque terms, one can say that the circle of the largest circumference that can be embedded within the torus does not have to be the outermost one. A circle located at the equatorial plane somewhere in the middle of the torus can have a larger circumference. This behavior is illustrated for our models in Fig.\ \ref{rc1}. The examples in Fig.\ \ref{rc1} have been chosen carefully---we plot the graphs corresponding to pairs of solutions with the same total mass $m_\mathrm{ADM}$. Each pair consist of a magnetized model and a model with no magnetic field. The differences between the graphs of $r_\mathrm{c}(r)$ in each pair are small (but visible on the plots). This is again consistent with the picture where the main influence the magnetic field has on the spacetime geometry is dynamical---through the distribution of the rest-mass density $\rho$ of the gas.

Another characteristic of strong spacetime curvature in black hole-disk systems is an occurrence of toroidal ergoregions. An ergoregion is defined as a region of the spacetime, outside the black hole horizon, in which the Killing vector $\xi^\mu$ (which is asymptotically timelike) becomes spacelike, i.e.,
\begin{equation}
g_{\mu\nu} \xi^\mu \xi^\nu = g_{tt} = - \alpha^2 + \psi^4 r^2 \sin^2 \theta \beta^2 > 0.
\end{equation}
The term ergosurface is sometimes reserved for the surface defined by a condition $g_{\mu \nu} \xi^\mu \xi^\nu = 0$.

A rotating Kerr black hole is surrounded by an ergoregion, usually called an ergosphere. In our case, rotating black holes are also surrounded by ergoregions (Fig.\ \ref{profrhoErgbp}). Examples of toroidal ergoregions associated with matter rotating around black holes can be found in \cite{ansorg, meinel, Herdeiro15, ergosfery}. Motivated by examples of toroidal ergoregions given in \cite{ansorg}, Chru\'{s}ciel, Greuel, Meinel, and Szybka obtained interesting mathematical results on the regularity of ergosurfaces in the vacuum region \cite{chrusciel}.

Toroidal ergoregions can be located within the torus, or conversely, the torus can be embedded within a toroidal ergoregion. For particularily compact systems, the ergoregions associated with the torus and with the black hole can merge. All these types of behavior were observed for non-magnetized black hole-torus systems in \cite{ansorg} and \cite{ergosfery}.

Examples of configurations with complex ergoregions are shown in in Figs.\ \ref{profrhoErgbp}--\ref{profrhoErgbp3}. In these plots ergoregions are marked in grey, while the contours of constant density are depicted with dotted lines. For comparison, we show configurations with the same parameters $m=1$, $a=0.9$, $\Gamma = 4/3$, $R_1 = 10$, $R_2 = 25$. The plots are grouped in pairs, showing configurations corresponding to the same total mass $m_\mathrm{ADM}$. The first configuration in each pair has been obtained assuming a non-zero magnetic field, while the second corresponds to an unmagnetized disk. Solutions with magnetic fields are computed assuming $n = 2$ and $C_1 = 0.5$. The changes in the shapes of the ergoregions are consistently small.

\section{Concluding remarks}

We have studied black hole-magnetized disk configurations, taking into account self-gravity of the disk. The survey of light disk configurations with different magnetization prescriptions yields results that are consistent with initial findings of \cite{magnetic_nasze}. This applies to the location of the ISCO, the behavior of the magnetic and thermal pressures, and their relation to the rest-mass density.

Allowing for larger disk masses, we recover a bifurcation pattern known from Ref.\ \cite{ergosfery} also for magnetized configurations. The general picture emerging from this part of our study is that the geometric effects---the growth of the proper volume of the torus, non-monotonicity of the circumferential radius, existence of toroidal ergoregions associated with the disks---are mainly caused by the distribution of the fluid component, which in turn can be affected by the magnetic field (cf.\ Figs.\ \ref{bif1dvol}, \ref{rc1}). In general, the magnetization of configurations with fixed parameters $n$ and $C_1$ in Eq.\ (\ref{magnetizationlaw}) decreases with the total mass $m_\mathrm{ADM}$ (the magnetization parameter $\beta_\mathrm{mag}$ increases with $m_\mathrm{ADM}$).

The existence of black hole-disk configurations with disconnected ergoregions, observed in \cite{labranche, meinel, ergosfery}, is now confirmed also for models with the magnetic field. This suggests a potential application in the context of the Blandford-Znajek effect \cite{blandford}.

Solutions corresponding to massive disks are, most likely, unstable. For non-magnetized, differentially rotating fluids there is a simple necessary condition for linear stability due to Seguin \cite{seguin}. An application of Seguin's criterion to solutions derived in \cite{ergosfery} suggests that the solutions corresponding to massive disks should be dynamically unstable. On the other hand solutions corresponding to light disks investigated in \cite{ergosfery} satisfy Seguin's condition. The stability analysis of magentized rotating fluids is, of course, much more involved, and we have to postpone it to another paper.

Our restriction to toroidal magnetic fields constitutes a strong simplification, both from the physical and technical point of view. On the other hand, configurations with toroidal magnetic fields seem to be physically relevant. For instance, numerical simulations of binary neutron star mergers suggest an occurrence of a post merger remanant consisting  of a black hole surrounded by a compact, quasi-stationary torus equipped with a mostly toroidal magnetic field \cite{kawamura}.

\section*{Acknowledgments}
We would like to thank Andrzej Odrzywo{\l}ek and Wojciech Kulczycki for their input in the optimization of the numerical code used in this paper. We also like to thank Edward Malec for careful reading of the manuscript of this paper. P.\ M.\ was partially supported by the Polish National Science Centre Grant No.\ 2017/26/A/ST2/00530.

\end{document}